\documentclass[3p,times]{elsarticle}

\usepackage{amssymb}
\usepackage{amsmath, bbm}
\usepackage{multirow}
\usepackage{longtable}
\usepackage[figuresright]{rotating}
\usepackage{float}
\usepackage{graphics}
\usepackage{psfrag}
\usepackage{epsfig}
\usepackage{epsf}

\newcommand{\be}{\begin{equation}}
\newcommand{\ee}{\end{equation}}

\newcommand{\beqa}{\begin{eqnarray}}
\newcommand{\eeqa}{\end{eqnarray}}

\newcommand{\fet}[1]{\mbox{\boldmath $#1$}}

\newcommand{\beq}{\begin{equation}}
\newcommand{\eeq}{\end{equation}}
\newcommand{\nn}{\nonumber \\ }

\def\bea{\arraycolsep .1em \begin{eqnarray}}
\def\eea{\end{eqnarray}}

\newlength{\feynwidth} 
\newlength{\feynwidthbig} 

\usepackage{xcolor} 

\allowdisplaybreaks

\begin{document}

\begin{frontmatter}


\title{Antinucleon-nucleon interaction at next-to-next-to-next-to-leading order\\
in chiral effective field theory}
\author[J]{Ling-Yun Dai}
\author[J]{Johann Haidenbauer}
\author[B,J]{Ulf-G. Mei{\ss}ner}
\address[J]{Institut f\"ur Kernphysik, Institute for Advanced Simulation and
J\"ulich Center for Hadron Physics,\\ Forschungszentrum J{\"u}lich, D-52425 J{\"u}lich, Germany}
\address[B]{Helmholtz Institut f\"ur Strahlen- und Kernphysik and Bethe Center
 for Theoretical Physics, Universit\"at Bonn, D-53115 Bonn, Germany}
\begin{abstract}
Results for the antinucleon-nucleon ($\bar NN$) interaction obtained at
next-to-next-to-next-to-leading order in chiral effective field theory (EFT) are reported.
A new local regularization scheme is used for the pion-exchange contributions
that has been recently suggested and applied in a pertinent study
of the $NN$ force within chiral EFT.
Furthermore, an alternative strategy for estimating the uncertainty is utilized
that no longer depends on a variation of the cutoffs.
The low-energy constants associated with the arising contact terms are fixed
by a fit to the phase shifts and inelasticities provided by a phase-shift analysis
of $\bar pp$ scattering data.
An excellent description of the $\bar NN$ amplitudes is achieved at the
highest order considered. Moreover, because of the quantitative reproduction of
partial waves up to $J=3$, there is also a nice agreement on the level of $\bar pp$
observables. Specifically, total and integrated elastic and charge-exchange
cross sections agree well with the results from the partial-wave
analysis up to laboratory energies of $300$~MeV, while differential
cross sections and analyzing powers are described quantitatively up
to $200$-$250$~MeV. The low-energy structure of the $\bar NN$ amplitudes
is also considered and compared to data from antiprotonic hydrogen.
\end{abstract}
\end{frontmatter}

\section{Introduction}
\label{sec:1}

The Low Energy Antiproton Ring (LEAR) at CERN has provided
a wealth of data on antiproton-proton ($\bar pp$) scattering
\cite{Rev1,Rev2,Rev3} and triggered a great number of pertinent
investigations~\cite{Dover80,Dover82,Cote,Timmers,Hippchen,Mulla,Timmermans,Mull}.
Its closure in 1996 has led to a noticeable
quiescence in the field of low-energy antiproton physics.
However, over the last decade there has been a renewed interest in
antinucleon-nucleon ($\bar NN$) scattering phenomena, prompted
for the main part by measurements of the $\bar pp$ invariant mass
in the decays of heavy mesons such as $J/\psi$,
$\psi'$, and $B$, and of the reaction cross section for $e^+e^- \to \bar pp$.
In several of those reactions a near-threshold enhancement in
the mass spectrum was found \cite{Bai,Aubert,Aubert1,BES12}.
While those observations nourished speculations about new resonances,
$\bar pp$ bound states, or even more exotic objects in some parts of
the physics community,
others noted that such data could provide a unique opportunity
to test the $\bar pp$ interaction at very low energies
\cite{BuggFSI,Zou,Sibirtsev05,Loiseau,JH06,JH06a,Entem,Dedonder,JH12,Haidenbauer:2014,Kang:2015,Dmitriev:2015,Milstein:2017}.
Indeed, in the aforementioned decays one has access to information on
$\bar pp$ scattering at significantly lower energies than it was
ever possible at LEAR.
In the future one expects a further boost of activities related to the $\bar NN$
interaction due to the Facility for Antiproton and Ion Research (FAIR)
in Darmstadt whose construction is finally on its way \cite{PANDA}.
In the course of this renewed interest new phenomenological $\bar NN$ potential
models have been published~\cite{Entem06,Bennich}. Moreover,
an update of the Nijmegen partial-wave analysis (PWA) of antiproton-proton
scattering data \cite{Timmermans} has been presented \cite{Zhou:2012}.

Over the same time period another important developement took place,
namely the emergence of chiral effective field theory (EFT)
as a powerful tool for the derivation of nuclear forces.
This approach, suggested by Weinberg \cite{Wei90,Wei91} and first put
into practice by van~Kolck and collaborators \cite{Ordonez:1993}, is now
at a stage where it facilitates a rather accurate and consistent
description of the $NN$ interaction and nuclear few-body systems, as
demonstrated in several publications, see e.g.~\cite{Entem:2003ft,Epe05,EKM:2015}.
Its most salient feature is that there is an underlying power counting which
allows one to improve calculations systematically by going to higher orders
in a perturbative expansion. With regard to the $NN$ force the
corresponding chiral potential contains pion exchanges and a series of
contact interactions with an increasing number of derivatives. The latter
represent the short-range part of the $NN$ force and are parameterized by
low-energy constants (LECs), that need to be fixed by a fit to data.
The reaction amplitude is obtained from solving a regularized
Lippmann-Schwinger equation for the derived interaction potential.
For an overview we refer the reader
to recent reviews~\cite{Epelbaum:2008ga,Machleidt:2011zz}.
A pedagogical introduction to the main concepts is given in~\cite{Epelbaum:2012vx}.

The $\bar NN$ interaction is closely connected to that in the $NN$ system
via $G$-parity. Specifically, the $G$-parity transformation (a combination
of charge conjugation and a rotation in the isospin space) relates
that part of the $\bar NN$ potential which is due to
pion exchanges to the one in the $NN$ case in an unambiguous way.
Thus, like in the $NN$ case, the long-range part of the $\bar NN$
potential is completely fixed by the underlying chiral symmetry
of pion-nucleon dynamics. Indeed, this
feature has been already exploited in the new PWA of
Ref.~\cite{Zhou:2012}. In this potential-based analysis the
long-range part of the utilized $\bar NN$ interaction consists of
one-pion exchange and two-pion-exchange contributions derived within
chiral EFT.

In this paper we present a $\bar NN$ potential derived in a chiral EFT
approach up to next-to-next-to-next-to leading order (N$^3$LO).
Its evaluation is done in complete analogy to the $NN$ interaction
published in Ref.~\cite{EKM:2015} and based on a modified Weinberg
power counting employed in that work.
In Ref.~\cite{Kang:2013} we had already studied the $\bar NN$ force
within chiral EFT up to next-to-next-to leading order (N$^2$LO).
It had been found that the approach works very well. Indeed, the overall
quality of the description of the $\bar NN$ amplitudes achieved in
Ref.~\cite{Kang:2013} is comparable to the one found in case of the $NN$
interaction at the same order \cite{Epe04a}.
By going to a higher order we expect that we will be able to
describe the $\bar NN$ interaction over a larger energy range.
Specifically, at N$^3$LO contact terms with four derivatives arise.
Consequently, now there are also low-energy constants that
contribute to the $D$ waves and can be used to improve the description of the
corresponding phase shifts.

Another motivation for our work comes from new developments in the treatment of the $NN$
interaction within chiral EFT. The investigation presented in Ref.~\cite{EKM:2015}
suggests that the nonlocal momentum-space regulator employed in the
$NN$ potentials in the past~\cite{Epe04a,Epe05}, but also in our application
to $\bar NN$ scattering~\cite{Kang:2013}, is not the most efficient choice,
since it affects the long-range part of the interaction.
In view of that a new regularization scheme that is defined in coordinate
space and, therefore, local has been proposed there. We adopt this scheme
also for the present work. After all, according to \cite{EKM:2015,Binder:2015}
this new regularization scheme does not distort the low-energy analytic structure
of the partial-wave amplitudes and, thus, allows for a better description of the
phase shifts.
Furthermore, in that work a simple approach for estimating
the uncertainty due to truncation of the chiral expansion is proposed
that does not rely on cutoff variation.  As shown in Ref.~\cite{Furnstahl:2015rha}
 this  procedure emerges generically from one class of Bayesian naturalness priors,
 and that all such priors result in consistent quantitative predictions for 68\% degree-of-believe
 intervals. We will adopt this approach for
performing an analogous analysis for our $\bar NN$ results.

Finally, at N$^3$LO it becomes sensible to compute not only phase shifts
but also observables and compare them directly with scattering data for
$\bar pp$ elastic scattering and for the charge-exchange reaction
$\bar pp \to \bar nn$. Such calculations have to be performed
in the particle basis because then the Coulomb interaction in the $\bar pp$
system can be taken into account rigorously as well as the different
physical thresholds of the $\bar pp$ and $\bar nn$ channels.

The present paper is structured as follows:
The elements of the chiral EFT $\bar NN$ potential up to N$^3$LO are
summarized in Section~2. Explicit expressions for the contributions
from the contact terms are given while those from pion exchange
are collected in \ref{app:potential}. The main emphasis in Section~2 is on
discussing how we treat the annihilation processes.
In this section we introduce also the Lippmann-Schwinger equation that
we solve and the parameterization of the S-matrix that we use.
In Section~3 we describe our fitting procedure. The LECs that arise
in chiral EFT, as mentioned above, are fixed by a
fit to the phase shifts and inelasticities provided by a recently
published phase-shift analysis of $\bar pp$ scattering data \cite{Zhou:2012}.
In addition we outline the procedure for the uncertainty analysis, which
is taken over from Ref.~\cite{EKM:2015}.
Results achieved up to N$^3$LO are presented in Section~4.
Phase shifts and inelasticity parameters for $S$, $P$, $D$,
and $F$ waves, obtained from our EFT interaction, are displayed
and compared with those of the $\bar NN$ phase-shift analysis.
Furthermore, results for various $\bar pp \to \bar pp$ and
$\bar pp \to \bar nn$ observables are given.
Finally, in Section~5, we analyze the low-energy structure of
the $N\bar N$ amplitudes and provide
predictions for $S$- and $P$-wave scattering lengths (volumes).
We also consider $\bar n p$ scattering.
A summary of our work is given in Section~6.
The explicit values of the four-nucleon LECs for the various fits are
tabulated in  \ref{app:LECs}.


\section{Chiral potential at next-to-next-to-next-to-leading order}
\label{sec:2}

\begin{figure}[t!]
\vskip 0.5cm
\includegraphics[width=\textwidth,keepaspectratio,angle=0,clip]{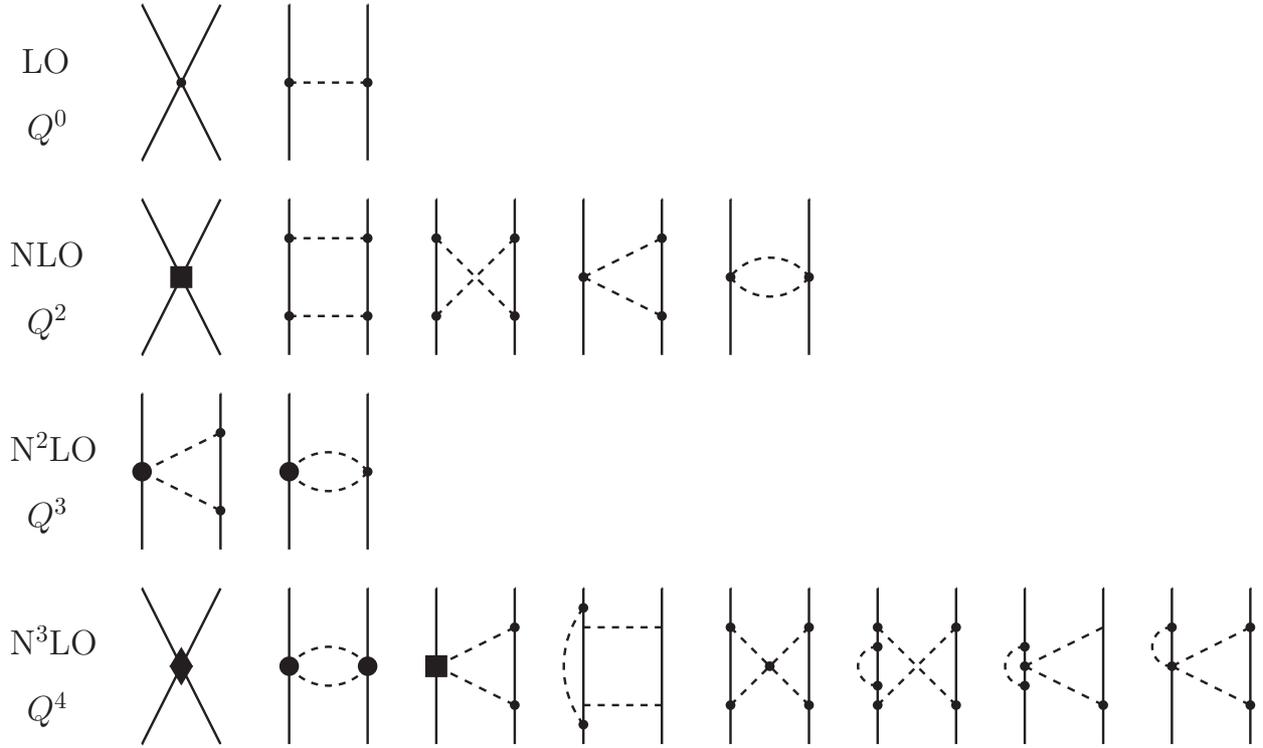}
\caption{Relevant diagrams up-to-and-including N$^3$LO.
Solid and dashed lines denote antinucleons/nucleons and pions, respectively.
The square and diamond symbolize contact vertices with two and four derivatives, respectively.
The dot denotes a leading $\pi N$ vertex while the filled circle denotes a subleading $\pi N$ vertex.
$Q$ denotes a small parameter (external momentum  and/or pion mass).
}
\label{fig:feynman}
\end{figure}

In chiral EFT the potential is expanded in powers of a quantity $Q = \tilde q / \Lambda_b$
in accordance with the employed power-counting scheme.
Here, $\tilde q$ stands for a soft scale that is associated with the typical momenta
of the nucleons or the pion mass and $\Lambda_b$ refers to the hard scale,
i.e. to momenta where the chiral EFT expansion is expected to break down.
The latter is usually assumed to be in the order of the rho mass.
The chiral potential up to N$^3$LO consists of contributions from
one-, two-, and three-pion exchange and of contact terms with up to four
derivatives \cite{EKM:2015}.
For a diagrammatic representation see Fig.~\ref{fig:feynman}.
Since the structure of the $\bar NN$ interaction is practically identical to the
one for $NN$ scattering, the potential given in Ref.~\cite{EKM:2015} can be
adapted straightforwardly for the $\bar NN$ case. However,
for the ease of the reader and also for defining our potential uniquely we
summarize the essential features below and we also provide explicit
expressions in \ref{app:potential}.

\subsection{Pion-exchange contributions}
\label{sec:2MEX}

The one-pion exchange potential is given by
\begin{equation}
\label{opep_full}
V_{1\pi} (q) =  \biggl(\frac{g_A}{2F_\pi}\biggr)^2 \,
\left( 1 - \frac{p^2 + p'^2}{2 m^2} \right)
\mbox{\boldmath $\tau$}_1 \cdot \mbox{\boldmath $\tau$}_2 \,
\frac{\mbox{\boldmath $\sigma$}_1 \cdot{\bf q}\,\mbox{\boldmath$\sigma$}_2\cdot{\bf q}}
{{\bf q}^2 + M_\pi^2} \ ,
\end{equation}
where ${\bf q}={\bf p}'-{\bf p}$ is the transferred momentum defined in terms of the
final (${\bf p}'$) and initial (${\bf p}$) center-of-mass momenta of the baryons
(nucleon or antinucleon). $M_\pi$ and $m$ denote the pion and antinucleon/nucleon
mass, respectively. Following \cite{Kang:2013} relativistic $1/m^2$ corrections
to the static one-pion exchange potential are taken into account already at NLO.
As in the work \cite{EKM:2015}
we take the larger value $g_A = 1.29$ instead of $g_A = 1.26$ in order to
account for the Goldberger--Treiman discrepancy.  This value, together with
the used $F_\pi = 92.4$ MeV, implies the pion-nucleon coupling
constant $g_{NN\pi} =13.1$ which is consistent with the empirical
value obtained from $\pi N$ and $NN$ data \cite{deSwart,Bugg} and also
with modern determinations utilizing the GMO sum rule \cite{Baru:2011bw}.
Contrary to \cite{EKM:2015}, isospin-breaking in the hadronic interaction
due to different pion masses is not taken into account. Here
we use the isospin-averaged value $M_\pi=138.039$~MeV.
The calculation of the $\bar NN$ phase shifts is done in the isospin
basis and here we adopt the average nucleon value $m=938.918$~MeV.
However, in the calculation of observables the mass difference between
protons and neutrons is taken into account and the corresponding values
from the PDG \cite{PDG} are used.

Note that the contribution of
one-pion exchange to the $\bar NN$ interaction is of opposite sign as that in
$NN$ scattering. This sign difference arises from the G-parity transformation of
the $NN\pi$ vertex to the $\bar N \bar N\pi$ vertex.
The contributions from two-pion exchange to $NN$ and $\bar NN$ are
identical. There would be again a sign differences for three-pion exchange.
However, since the corresponding contributions are known to be weak
we neglect them here as it was done in the $NN$ case \cite{Epelbaum:2014sza}.

The underlying effective pion-nucleon Lagrangian is given in Ref.~\cite{Fettes:2000gb}.
For the LECs $c_{i}$ and $\bar d_{i}$ that appear in the subleading $\pi\pi NN$
vertices we take the same values as in Ref.~\cite{EKM:2015}. Specifically,
for $c_{1}$, $c_{3}$, and $c_{4}$ we adopt the central values
from the $Q^3$--analysis of the $\pi N$ system \cite{Paul}, i.e.
$c_1=-0.81$ GeV$^{-1}$, $c_3=-4.69$~GeV$^{-1}$, $c_4=3.40$ GeV$^{-1}$,
while $c_2= 3.28$ GeV$^{-1}$ is taken from the heavy-baryon calculation in
Ref.~\cite{Fettes:1998}. However, in the future the more precise values of the
$c_i$ determined from the Roy-Steiner analysis of pion-nucleon scattering~\cite{Hoferichter:2015tha}
should be used for the $NN$ as well as the $\bar NN$ case.
Note also that different values for the $c_{i}$ were used in the
$\bar NN$ PWA \cite{Zhou:2012}. Therefore, the two-pion exchange potential employed
in our analysis differs from the one used for determining the $\bar NN$ phase
shifts. However, based on the uncertainty estimate given in Ref.~\cite{Zhou:2012}
we do not expect any noticeable effects from that on the quality of our results.
In any case, it has to be said that our calculation includes also N$^3$LO
corrections to the two-pion exchange so that the corresponding potentials
differ anyway.

In this context let us mention another difference to the analysis in Ref.~\cite{Zhou:2012}.
It concerns the electromagnetic interaction where we consider only the
(non-relativistic) Coulomb interaction in the $\bar pp$ system, but we
neglect the magnetic-moment interaction.

\subsection{Contact terms}
\label{sec:2CT}

The contact terms in partial-wave projected form are given by \cite{EKM:2015}
\begin{eqnarray}
\label{VC0}
V(^1S_0)
&=& \tilde{C}_{^1S_0} + {C}_{^1S_0} ({p}^2+{p}'^2)
+{D^1}_{^1S_0}{p}^2{p}'^2 + {D^2}_{^1S_0} ({p}^4+{p}'^4)\,, \label{C1S0}\\
V(^3S_1)
&=& \tilde{C}_{^3S_1} + {C}_{^3S_1} ({p}^2+{p}'^2)
+{D^1}_{^3S_1}{p}^2{p}'^2 + {D^2}_{^3S_1} ({p}^4+{p}'^4)\,, \label{C3S1}\\
V(^1P_1)
&=& {C}_{^1P_1}\, {p}\, {p}'+{D}_{^1P_1}\, {p}\, {p}'({p}^2+{p}'^2)\,,\\
V(^3P_1)
&=& {C}_{^3P_1}\, {p}\, {p}'+{D}_{^3P_1}\, {p}\, {p}'({p}^2+{p}'^2)\,, \\
V(^3P_0)
&=& {C}_{^3P_0}\, {p}\, {p}'+{D}_{^3P_0}\, {p}\, {p}'({p}^2+{p}'^2)\,,\\
V(^3P_2)
&=& {C}_{^3P_2}\,  {p}\, {p}'+{D}_{^3P_2}\, {p}\, {p}'({p}^2+{p}'^2)\,,\\
V(^3D_1 -\, ^3S_1)
&=& {C}_{\epsilon_1}\, {p'}^2+{D^1}_{\epsilon_1}{p}^2{p}'^2 + {D^2}_{\epsilon_1} {p}'^4 \,,\\
V(^3S_1 -\, ^3D_1)
&=& {C}_{\epsilon_1}\, {p}^2+{D^1}_{\epsilon_1}{p}^2{p}'^2 + {D^2}_{\epsilon_1} {p}^4 \,,\\
V(^3D_1)
&=& {D}_{^3D_1}\, {p}^2 {p}'^2\,,\\
V(^1D_2)
&=& {D}_{^1D_2}\, {p}^2 {p}'^2\,,\\
V(^3D_2)
&=& {D}_{^3D_2}\, {p}^2 {p}'^2\,,\\
V(^3F_2 -\, ^3P_2)
&=& {D}_{\epsilon_2}{p}{p}'^3\,,\\
V(^3P_2 -\, ^3F_2)
&=& {D}_{\epsilon_2}{p}^3{p}'\,,\\
\label{VC}
\end{eqnarray}
with $p = |{\bf p}\,|$ and ${p}' = |{\bf p}\,'|$.
Here, the $\tilde{C}_i$ denote the LECs that arise at LO and that correspond to
contact terms without derivates, the ${C}_i$ arise at NLO  from
contact terms with two derivates, and ${D}_i$ are those at N$^3$LO
 from contact terms with four derivates.
Note that the Pauli principle is absent in case of the $\bar NN$ interaction.
Accordingly, each partial wave that is allowed by angular momentum conservation
occurs in the isospin $I=0$ and in the $I=1$ channel.
Therefore, there are now twice as many contact terms as in $NN$, that means
$48$ up to N$^3$LO.

The main difference between the $NN$ and $\bar NN$ interactions
is the presence of annihilation processes in the latter. Since the
total baryon number is zero, the $\bar NN$ system can annihilate
and this proceeds via a decay into multi-pion channels, where typically
annihilation into 4 to 6 pions is dominant in the low-energy region of
$\bar NN$ scattering \cite{Rev1}.

Since annihilation is a short-ranged process as argued in Ref.~\cite{Kang:2013},
in principle, it could be taken into account by simply using complex LECs in
Eqs.~(\ref{VC0})-(\ref{VC}). Indeed, this has been done in some EFT studies
of $\bar NN$ scattering~\cite{Chen2010,Chen2011}.
However, with such an ansatz it is impossible to impose sensible unitarity conditions.
Specifically, there is no guarantee that the resulting scattering amplitude fulfills
the optical theorem, i.e. a requirement which ensures that for each partial wave
the contribution to the total cross section is larger than its contribution to
the integrated elastic cross section.
Therefore, in Ref.~\cite{Kang:2013} we treated annihilation in a different way
so that unitarity is manifestly fulfilled already on a formal level.
It consisted in considering the annihilation potential to be due to an effective
two-body annihilation channel $X$ for each partial wave,
\begin{equation}
V_{ann} = V_{\bar NN \to X} G_X^{} V_{X\to \bar NN} ,
\label{ANN}
\end{equation}
with $V_{\bar NN \to X}$ the transition potential.
Under the assumption that the threshold of $X$ is significantly below the one of
$\bar NN$ the center-of-mass momentum in the annihilation channel is already fairly
large and its variation in the low-energy region of $\bar NN$
scattering considered here can be neglected. Then the transition
potential $V_{\bar NN \to X}$ can be represented by contact terms similar to
the ones for $\bar NN \to \bar NN$, cf. Eqs.~(\ref{VC0})-(\ref{VC}), and the
Green's function $G_X^{}$ reduces to the unitarity cut, i.e. $G_X^{} \propto -{\rm i}$.
Note that Eq.~(\ref{ANN}) is exact under the assumption that there is no
interaction in and no transition between the various annihilation channels.

The annihilation part of the $\bar NN$ potential is then of the form
\begin{eqnarray}
V^{L=0}_{ann} &=& -i\, (\tilde C_{^1S_0}^a+C_{^1S_0}^ap^2+D_{^1S_0}^ap^4)\,(\tilde C_{^1S_0}^a+C_{^1S_0}^ap'^2+D_{^1S_0}^ap'^4), \\
V^{L=1}_{ann} &=& -i\, (C_\alpha^a p+ D_\alpha^a p^3)\,(C_\alpha^a p'+ D_\alpha^a p'^3),\\
V^{L=2}_{ann} &=& -i\, (D_\beta^a)^2 p^2 p'^2,\\
V^{L=3}_{ann} &=& -i\, (D_\gamma^a)^2 p^3 p'^3,
\label{ANN1}
\end{eqnarray}
where $\alpha$ denotes the $^3P_0$, $^1P_1$, and $^3P_1$ partial waves,
$\beta$ stands for $^1D_2$, $^3D_2$ and $^3D_3$, and $\gamma$ for $^1F_3$, $^3F_3$ and $^3F_4$.
The superscript $a$ is used to distinguish the LECs from those in the elastic part of
the $\bar NN$ potential.
For the coupled $^3S_1-$$^3D_1$ partial wave we use
\begin{eqnarray}
V^{S\to S}_{ann} &=& -i\, (\tilde C_{^3S_1}^a+C_{^3S_1}^ap^2+D_{^3S_1}^ap^4)\,(\tilde C_{^3S_1}^a+C_{^3S_1}^ap'^2+D_{^3S_1}^ap'^4), \nonumber\\
V^{S\to D}_{ann} &=& -i\, (\tilde C_{^3S_1}^a+C_{^3S_1}^ap^2D_{^3S_1}^ap^4)\, C_{\epsilon_1}^a p'^2, \nonumber \\
V^{D\to S}_{ann} &=& -i\,  C_{\epsilon_1}^a p^2\, (\tilde C_{^3S_1}^a+C_{^3S_1}^ap'^2+D_{^3S_1}^ap'^4), \nonumber \\
V^{D\to D}_{ann} &=& -i\,  [(C_{\epsilon_1}^a)^2+(C_{^3D_1}^a)^2 ] p^2 p'^2 \ .
\label{ANN2}
\end{eqnarray}
and for $^3P_2-$$^3F_2$
\begin{eqnarray}
V^{P\to P}_{ann} &=& -i\, (C_{^3P_2}^a p+ D_{^3P_2}^a p^3)\,(C_{^3P_2}^a p'+ D_{^3P_2}^a p'^3),\nonumber\\
V^{P\to F}_{ann} &=& -i\, (C_{^3P_2}^a p+ D_{^3P_2}^a p^3)\,D_{\epsilon_2}^a p'^3,\nonumber\\
V^{F\to P}_{ann} &=& -i\,  D_{\epsilon_2}^a p^3(C_{^3P_2}^a p'+ D_{^3P_2}^a p'^3), \nonumber\\
V^{F\to F}_{ann} &=& -i\, [(D_{\epsilon_2}^a)^2+(D_{^3F_2}^a)^2] p^3 p'^3 \ .
\label{ANN3}
\end{eqnarray}
In the expressions above the parameters $\tilde C^a$, $C^a$, and $D^a$ are real. There is no
restriction on the signs of $\tilde C^a$, $C^a$, $D^a$ because the sign of $V_{ann}$ as required
by unitarity is already explicitly fixed. Note, however, that terms of the form $p^ip'^j$ with
higher powers $n=i+j$ than what follows from the standard Weinberg power counting arise in
various partial waves from unitarity constraints and those have to be included in order to
make sure that unitarity is fulfilled at any energy.
Still we essentially recover the structure of the potential that follows from the standard
power counting for $\bar NN\to \bar NN$ (cf. Eqs.~(\ref{VC0})-(\ref{VC})) with a similar (or
even identical) number of counter terms (free parameters) for the annihilation part.

As one can see in Eq.~(\ref{ANN1}) and also in Eq.~(\ref{ANN3}) we allowed for contact terms
in the annihilation potential for $F$ waves. This is motivated by two reasons.
First, according to the PWA there is a nonzero contribution of $F$ waves to the annihilation
cross section and we wanted to be able to take this into account.
Second, as can be seen in Eq.~(\ref{ANN3}), terms proportional to $p^3 p'^3$ appear anyway
in the $^3F_2$ partial wave because of unitarity constraints. Moreover, transitions
proportional to $p^3p'$ (for $^3F_2 \to$$^3P_2)$ are present in the real part at N$^3$LO,
see Eq.~(\ref{VC}). This suggests that the analogous type of transitions should be taken
into account in the description of annihilation via Eq.~(\ref{ANN}) from $F$ waves,
i.e. $V^F_{\bar NN \to X} \equiv D^a_F p^3$.
With regard to the real part of the $\bar NN$ (or $NN$)
potential contact terms proportional to $p^3 p'^3$ would first appear at N$^5$LO
in the standard Weinberg counting and here we do not depart from the counting.

Note that, in principle, there is a contribution from the principal-value
part of the integral in Eq.~(\ref{ANN}). However, it is real and, therefore,
its structure is already accounted for by the standard LECs in
Eqs.~(\ref{VC0})--(\ref{VC}).

\subsection{Scattering equation}\label{sec:LSE}

As first step a partial-wave projection of the interaction potentials is performed,
following the procedure described in detail in Ref.~\cite{Epe05}.
Then the reaction amplitudes are obtained from the solution of a relativistic
Lippmann-Schwinger (LS) equation:
\begin{eqnarray}
T_{L''L'}(p'',p';E_k)=V_{L''L'}(p'',p')+
\sum_{L}\int_0^\infty \frac{dpp^2}{(2\pi)^3} \, V_{L''L}(p'',p)
\frac{1}{2E_k-2E_p+i0^+}T_{LL'}(p,p';E_k).\nonumber\\
\label{LS}
\end{eqnarray}
Here, $E_k=\sqrt{m^2+k^2}$, where $k$ is the on-shell momentum.
We adopt a relativistic scattering equation so that our amplitudes
fulfill the relativistic unitarity condition at any order, as done also
in the $NN$ sector \cite{Epe05,Machleidt:2011zz}. On the other
hand, relativistic corrections to the potential are calculated order by
order. They appear first at next-to-next-to-next-to-leading order
(N$^3$LO) in the Weinberg scheme, see \ref{app:potential}.

Analogous to the $NN$ case we have either uncoupled spin-singlet and triplet
waves (where $L''=L'=L=J$) or coupled partial waves (where $L'',L',L=J-1,J+1$).
The LECs of the $\bar NN$ potential are determined by a fit to the phase shifts
and inelasticity parameters of Ref.~\cite{Zhou:2012}. Those quantities were
obtained under the assumption of isospin symmetry and, accordingly,
we solve the LS equation in the isospin basis where the $I=0$ and $I=1$
channels are decoupled.
For the calculation of observables, specifically for the direct comparison of our
results with data, we solve the LS equation in particle basis. In this case
there is a coupling between the $\bar pp$ and $\bar nn$ channels. The corresponding
potentials are given by linear combinations of the ones in the isospin basis,
i.e. $V^{\bar pp} = V^{\bar nn} = (V^{I=0}+V^{I=1})/2$ and
$V^{\bar pp\leftrightarrow\bar nn} = (V^{I=0}-V^{I=1})/2$.
Note that the solution of the LS equation in particle basis no longer fulfills isospin symmetry.
Due to the mass difference between $p$ ($\bar p$) and $n$ ($\bar n$) the physical
thresholds of the $\bar pp$ and $\bar nn$ channels are separated by
about 2.5~MeV. In addition the Coulomb interaction is present in the $\bar pp$ channel.
Both effects are included in our calculation where the latter is implemented via the
Vincent-Phatak method~\cite{VP}.
Other electromagnetic effects like those of the magnetic-moment interaction,
considered in Ref.~\cite{Zhou:2012} are, however, not taken into account in our calculation.

The relation between the  $S$-- and on--the--energy shell $T$--matrix is given by
\begin{equation}
\label{Sdef}
S_{L L'} (k) = \delta_{L L'} - \frac{i}{8 \pi^2}
\, k \, E_k \,  T_{L L'}(k)~.
\end{equation}
The phase shifts in the uncoupled cases can be obtained from the
$S$--matrix via
\begin{equation}
S_{LL} \equiv S_{L} = e^{ 2 i \delta_{L}} \, .
\label{SM0}
\end{equation}
For the $S$--matrix in the coupled channels ($J>0$) we
use the so--called Stapp parametrization~\cite{Stapp}
\beqa
\left( \begin{array}{cc} S_{J-1 \, J-1} &  S_{J-1 \, J+1} \\
S_{J+1 \, J-1} &  S_{J+1 \, J+1} \end{array} \right)
=
\left( \begin{array}{cc} \cos{2 \epsilon_J}\, e^{2 i \delta_{J-1}} &
-i \sin{2 \epsilon_J}\, e^{i(\delta_{J-1} + \delta_{J+1})} \\
-i \sin{2 \epsilon_J}\, e^{i(\delta_{J-1} + \delta_{J+1})} &
\cos{2 \epsilon_J}\, e^{2 i \delta_{J+1}} \end{array} \right)~.
\label{SM1}
\eeqa

In case of elastic scattering the phase parameters in Eqs.~(\ref{SM0})
and (\ref{SM1}) are real quantities while in the presence of inelasticities
they become complex. Because of that, in the past several generalizations
of these formulae have been proposed that still allow one to write the
$S$-matrix in terms of real parameters \cite{Arndt,Zhou:2012}.
We follow here Ref.~\cite{Bystricky} and calculate/present simply
the real and imaginary parts of the phase shifts and the mixing
parameters obtained via the above parameterization. Note that with this
choice the real part of the phase shifts is identical to the phase shifts
one obtains from another popular parameterization where the imaginary
part is written in terms of an inelasticity parameter $\eta$, e.g.
for uncoupled partial waves
\begin{equation}
S_{L} = \eta e^{ 2 i \delta_{L} } \; .
\label{SM2}
\end{equation}
Indeed, for this case ${\rm Im}\, \delta_{L} = -(\log \eta ) / 2$
which implies that ${\rm Im}\, \delta_{L} \geq 0$
since $\eta \leq 1$ because of unitarity. Note that for simplicity reasons,
in the discussion of the results below we will refer to the real part of the
phase shift as {\it phase shift} and to the imaginary part as
{\it inelasticity parameter}.
Since our calculation implements unitarity, the optical theorem
\begin{equation}
{\rm Im} \, a_{LL}(k) \geq k \, \sum_{L'} |a_{LL'}(k)|^2 \ ,
\label{OT}
\end{equation}
is fulfilled for each partial wave, where
$ a_{LL'}(k) = (S_{LL'} - \delta_{LL'}) /(2ik) =
- 1/(4\pi)^2 \cdot E_k \, T_{L L'}(k)$.

For the fitting procedure and for the comparison of our results
with those of Ref.~\cite{Zhou:2012} we reconstructed the $S$-matrix
based on the phase shifts listed in Tables~VIII-X of that paper
via the formulae presented in Sect. VII of that paper and then
converted them to our convention specified in Eqs.~(\ref{SM0}) and
(\ref{SM1}).

\section{Fitting procedure and estimation of the theoretical uncertainty}

An important objective of the work of Ref.~\cite{EKM:2015}
consisted in a careful analysis of the cutoff dependence and in providing
an estimation of the theoretical uncertainty. The reasoning for making
specific assumptions, and adopting and following specific procedures in order
to achieve that aim has been explained and thoroughly discussed in that paper
and we do not repeat this here in detail.
However, we want to emphasize that whatever has been said there for
$NN$ scattering is equally valid for the $\bar NN$ system. It is a consequence
of the fact that the general structure of the long-range part of the two
interactions is identical -- though the actual potential strengths in the
individual partial waves certainly differ.
Accordingly, the non-local exponential regulator employed in \cite{Epe05,Epe04a}
but also in our N$^2$LO study of $\bar NN$ scattering \cite{Kang:2013}
for the one- and two-pion exchange contributions will be replaced here
by the new regularization scheme described in Sect.~3 of \cite{EKM:2015}.
This scheme relies on a regulator that is defined in coordinate space and,
therefore, is local by construction.
As demonstrated in that reference, the use of a local regulator is superior at
higher energies and, moreover, produces a much smaller amount of artefacts
over the whole considered energy range.
The contact interactions are non-local anyway, cf. Eqs.~(\ref{VC0})-({\ref{VC}).
In this case we use again the standard nonlocal regulator of Gaussian type.
The explict form of the cutoff functions employed in the present study is
given by \begin{equation}
f(r) = \left[ 1 - {\rm exp}\left(-\frac{r^2}{R^2}\right)\right]^n, \quad
f(p',p) = {\rm exp}\left(-\frac{p'^m+p^m}{\Lambda^m}\right) \ .
\label{cutoff}
\end{equation}

For the cutoffs we consider the same range as in Ref.~\cite{EKM:2015},
i.e from $R=0.8$~fm to $R=1.2$~fm. The cutoff in momentum-space applied
to the contact interactions is fixed by the relation
$\Lambda = 2 R^{-1}$ so that the corresponding range is then
$\Lambda \simeq 500, ..., 300$ MeV. Following \cite{EKM:2015},
the exponent in the coordinate-space cutoff function is chosen to be $n=6$,
the one for the contact terms in momentum space to be $m=2$.

\subsection{Fitting procedure}

In the fitting procedure we follow very closely the strategy of
Ref.~\cite{EKM:2015} in their study of the $NN$ interaction.
The LECs are fixed from a fit to the $\bar NN$ phase shifts and
mixing parameters of Ref.~\cite{Zhou:2012} where we take into account their
results for
$p_{lab}\leq 300$~MeV/c ($T_{lab}\leq 50$~MeV) at LO,
$p_{lab}\leq 500$~MeV/c ($T_{lab}\leq 125$~MeV) at NLO and N$^2$LO, and
$p_{lab}\leq 600$~MeV/c ($T_{lab}\leq 175$~MeV) at N$^3$LO.
Exceptions are made in cases where the phase shifts (or inelasticity
parameters) exhibit a resonance-like behavior at the upper end
of the considered momentum interval. Then we extend or reduce the
energy range slightly in order to stabilize the results and avoid artefacts.

No uncertainties are given for the $\bar NN$ phase shifts and
inelasticity parameters of the PWA. Because of that we
adopt a constant and uniform value $\Delta$ for them for the evaluation of
the function to which the minimization procedure is applied. Thus,
the uncertainty is reduced simply to an overall normalization factor.
On top of that, additional weight factors are introduced in the fitting process
in a few cases where it turned out to be difficult to obtain stable results.
The $\tilde\chi^2$ values summarized in Table~\ref{tab:chi2} for orientation
are, however, all calculated with a universal $\Delta$ which was set to
$\Delta^2=0.1$. The tilde is used as a reminder that these are not genuine
chi-square values.
The actual $\tilde\chi^2$ function in the fitting procedure for each partial
wave is $|S_{LL'}-S^{PWA}_{LL'}|^2/\Delta^2$ where the $S$-matrix elements
$S^{PWA}_{LL'}$ are reconstructed from the phase shifts and
inelasticity parameters given in Tables~VIII-X of Ref.~\cite{Zhou:2012}.

Table~\ref{tab:chi2} reveals that the lowest values for the $\tilde\chi^2$ are achieved
for hard cutoffs, namely $R=0.8-0.9$~fm. This differs slightly from the $NN$ case where
somewhat softer values $R=0.9-1.0$~fm seem to be preferred. In both cases a strong
increase in the $\tilde\chi^2$ is observed for the softest cutoff radius considered,
i.e. for $R=1.2$~fm.
For the illustration of our results we will use, in general, the interaction with the
cutoff $R=0.9$~fm. That value was found to be the optimal cutoff choice in the $NN$
study \cite{EKM:2015}. Nominally, in terms of the $\tilde\chi^2$ value, $R=0.8$~fm
would be the optimal cutoff choice for $\bar NN$. But the differences in the quality
of the two fits are so small, see Table~\ref{tab:chi2}, that we do not attribute any
significance to them given that no proper chi-square can be calculated.
The numerical values of the LECs are compiled in Tables in \ref{app:LECs}.
\begin{table}[h!]
\begin{center}
\caption{\label{tab:chi2} Resulting effective $\tilde\chi^2$ (see text) for different cutoffs and different energy regions.}
 \vspace{0.3cm}
\renewcommand{\arraystretch}{1.2}
\begin{tabular}{c c c c c c}
\hline\hline
$\tilde\chi^2$  & \rule{0.2cm}{0cm} R=0.8~fm \rule{0.2cm}{0cm}      &  \rule{0.2cm}{0cm} R=0.9~fm \rule{0.2cm}{0cm}         & \rule{0.2cm}{0cm} R=1.0~fm \rule{0.2cm}{0cm}  & \rule{0.2cm}{0cm} R=1.1~fm \rule{0.2cm}{0cm} & \rule{0.2cm}{0cm} R=1.2~fm \rule{0.2cm}{0cm}     \\
\hline
$T_{lab}\leq 25$~MeV     & 0.003    &  0.004       &  0.004   &  0.019  &  0.036          \\
$T_{lab}\leq 100$~MeV    & 0.023    &  0.025       &  0.036   &  0.090  &  0.176          \\
$T_{lab}\leq 200$~MeV    & 0.106    &  0.115       &  0.177   &  0.312  &  0.626          \\
$T_{lab}\leq 300$~MeV    & 2.012    &  2.171       &  3.383   &  5.531  &  9.479          \\
\hline
\hline
\end{tabular}
\end{center}
\renewcommand{\arraystretch}{1.0}
\end{table}

\subsection{Estimation of the theoretical uncertainty}

The motivation and the strategy, and also the shortcomings, of the procedure
for estimating the theoretical uncertainty suggested in Ref.~\cite{EKM:2015}
are discussed in detail in Sect.~7 of that reference.
The guiding principle behind that suggestion is that one uses the expected
size of higher-order corrections for the estimation of the theoretical uncertainty.
This is commonly done, e.g.~in the Goldstone boson and single-baryon sectors
of chiral perturbation theory. This approach is anticipated to
provide a natural and more reliable estimate than relying on cutoff
variations, say, as done in the past, and, moreover, it has the advantage
that it can be applied for any fixed value of the cutoff $R$.

The concrete expression used in this approach to calculate
an uncertainty $\Delta X^{\rm N^3LO} (k)$ to the N$^3$LO prediction
$X^{\rm N^3LO}(k)$ of a given observable $X(k)$ is \cite{EKM:2015}
\begin{eqnarray}
\label{Error}
\Delta X^{\rm N^3LO} (k) &=& \max  \bigg( Q^5 \times \Big| X^{\rm
    LO}(k) \Big|, \;\;\;  Q^3 \times \Big|
  X^{\rm LO}(k) -   X^{\rm NLO}(k) \Big|, \;\; \;  Q^2 \times \Big|
  X^{\rm NLO}(k) -   X^{\rm N^2LO}(k) \Big|, \nonumber \\
&& {}  \;\;\;\;\;\; \; \; \; \;  Q \times \Big|
  X^{\rm N^2LO}(k) -   X^{\rm N^3LO}(k) \Big|  \bigg)\,,
\end{eqnarray}
where the expansion parameter $Q$ is defined by
\begin{equation}
\label{expansion}
Q =\max \left( \frac{k}{\Lambda_b}, \; \frac{M_\pi}{\Lambda_b} \right)\,,
\end{equation}
with $k$  the cms momentum corresponding to the considered
laboratory momentum and $\Lambda_b$ the breakdown scale.
For the latter we take over the values established in Ref.~\cite{EKM:2015}
which are $\Lambda_b=600\,$MeV for the cutoffs $R=0.8$, $0.9$ and
$1.0\,$fm, $\Lambda_b = 500\,$MeV for $R=1.1\,$fm
and $\Lambda_b=400\,$MeV for $R=1.2$.
Analogous definitions are used for calculating the uncertainty up to N$^2$LO,
etc.
Note that the quantity $X(k)$ represents not only a ``true'' observable such as
a differential cross section or an analyzing power, but also for a phase shift
or an inelasticity parameter.

As already emphasized in \cite{EKM:2015}, such a simple estimation of the
theoretical uncertainty does not provide a statistical interpretation.
Note, however, that this procedure can be interpreted in a Bayesian
sense~\cite{Furnstahl:2015rha}.
Let us also mention that -- like in \cite{EKM:2015} -- we
impose an additional constraint for the theoretical
uncertainties at NLO and N$^2$LO by requiring them to have at least
the size of the actual higher-order contributions.

\section{Results}

\subsection{Phase shifts}

%
Let us first consider the influence of cutoff variations on our results.
In Figs.~\ref{fig:ph-1}-\ref{fig:ph-3} phase shifts and inelasticity parameters
for partial waves up to a total angular momentum of $J=4$ are presented.
We use here the spectral notation $^{(2S+1)}L_J$ and indicate the isospin $I$
separately. Subscripts $R$ and $I$ are used for $\delta$ in order to distinguish
between the real and imaginary part of the phases and mixing angles.
The  cutoffs considered are $R=0.8$, $0.9$, $1.0$, $1.1$, and $1.2$~fm and the
results are based on the chiral potential up to N$^3$LO.

One can see that for most partial waves the cutoff dependence is fairly weak
for $T_{lab}$ up to $300$~MeV ($p_{lab}$ up to $800$~MeV/c). Indeed, the small
residual cutoff dependence that we observe here is comparable to the likewise
small variation reported in Ref.~\cite{EKM:2015} for the $NN$ interaction.
Only in a few cases there is a more pronounced cutoff dependence
of the results for energies above $150$-$200$~MeV. This has to do with the fact
that the PWA \cite{Zhou:2012} suggests a resonance-like behavior of some phases
in this region. This concerns most prominently the $^1S_0$ partial wave with isospin $I=1$
and the $^3P_0$ partial wave with $I=1$. In addition, also a few other
partial waves show a conspicuous behavior at higher energies in the sense
that the energy dependence changes noticeably. Typical examples are the
inelasticity parameters for the $I=0$ $^3P_0$ and $^3P_2$ partial wave,
where the corresponding $\delta_{\rm I}$'s increase rapidly from the threshold,
but then level out at higher energies. Describing this behavior with the two LECs
at N$^3$LO, that have to absorb the cutoff dependence at the same time, is
obviously only possible for a reduced energy region.

\begin{figure}[htbp]
\vspace{-2.0cm}
\centering
\includegraphics[width=1.0\textwidth,height=1.0\textheight]{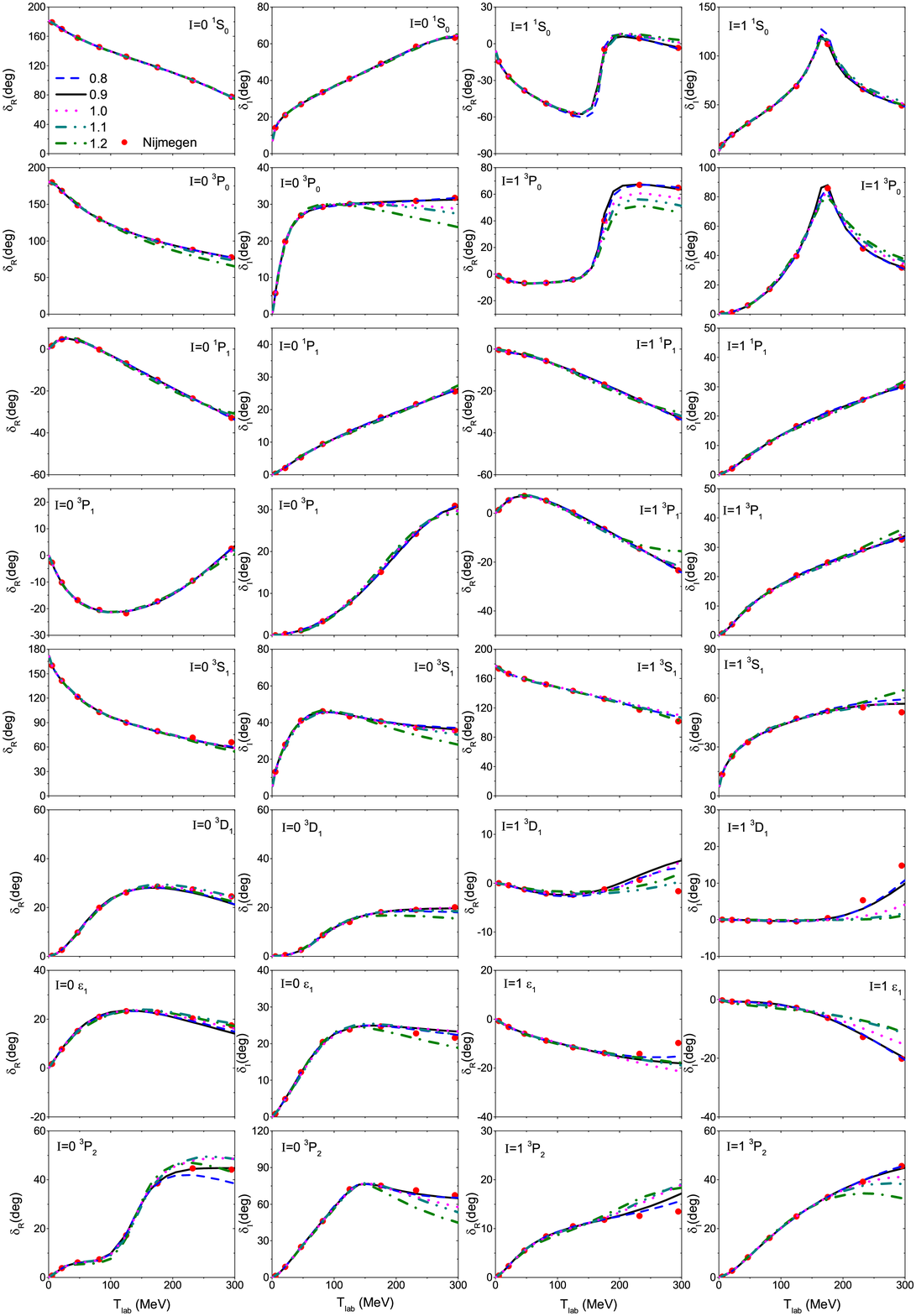}
\caption{Real and imaginary parts of various $\bar NN$ phase shifts
at N$^3$LO for cutoffs $R=0.8-1.2$~fm.
The filled circles represent the solution of the $\bar pp$ PWA~\cite{Zhou:2012}.
}
\label{fig:ph-1}
\end{figure}

\begin{figure}[htbp]
\vspace{-2.0cm}
\centering
\includegraphics[width=1.0\textwidth,height=1.0\textheight]{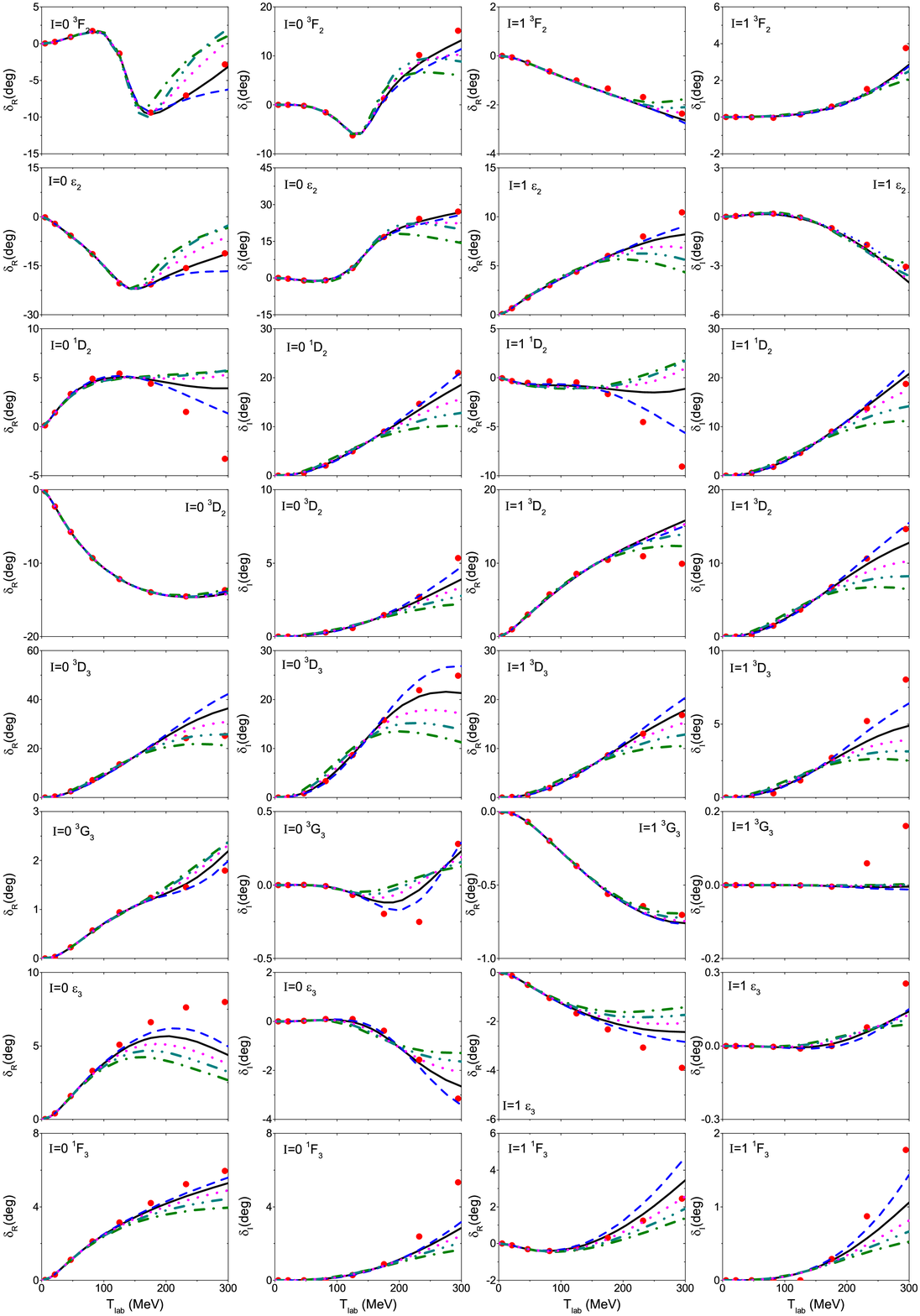}
\caption{Real and imaginary parts of various $\bar NN$ phase shifts
at N$^3$LO for cutoffs $R=0.8-1.2$~fm.
The filled circles represent the solution of the $\bar pp$ PWA~\cite{Zhou:2012}. \label{fig:ph-2}}
\end{figure}

\begin{figure}[htbp]
\vspace{-2.0cm}
\centering
\includegraphics[width=1.0\textwidth,height=0.5\textheight]{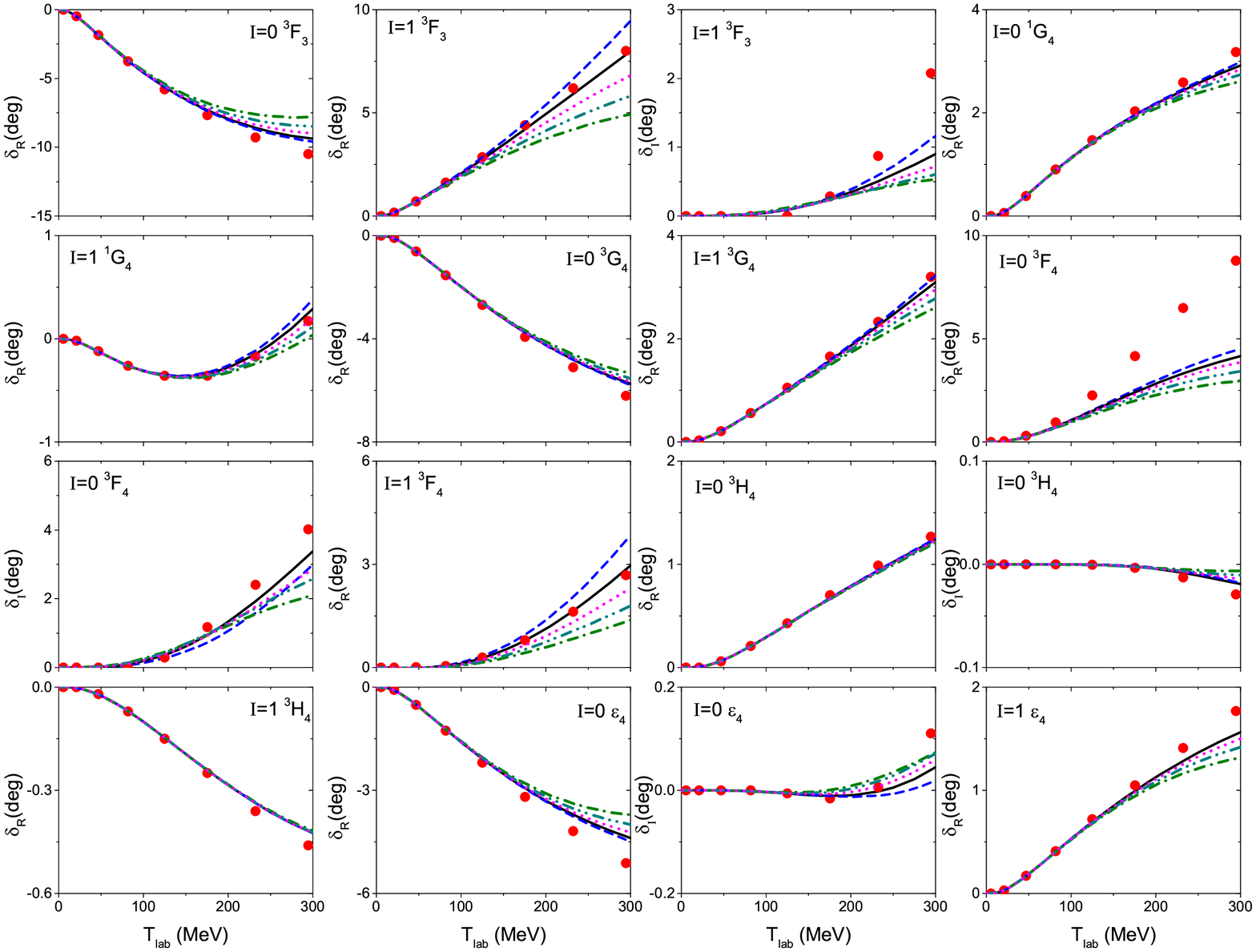}
\caption{Real and imaginary parts of various $\bar NN$ phase shifts
at N$^3$LO for cutoffs $R=0.8-1.2$~fm.
The filled circles represent the solution of the $\bar pp$ PWA~\cite{Zhou:2012}.
\label{fig:ph-3}}
\end{figure}

\begin{figure}[thbp]
\centering
\includegraphics[width=0.9\textwidth,height=0.45\textheight]{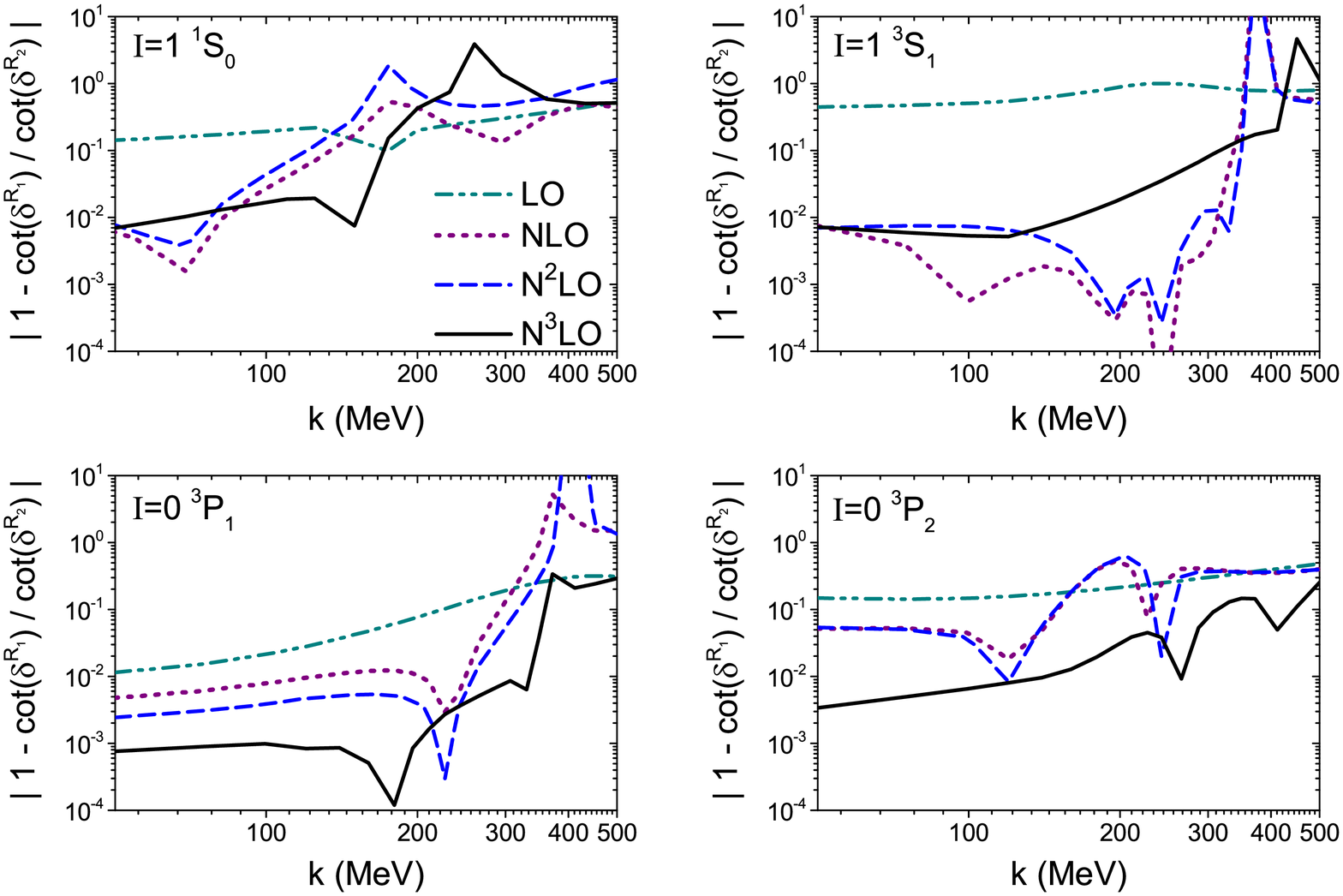}
\caption{Error plot for the real part of the phase shifts for the isospin $I=1$ partial waves
$^1S_0$, $^3S_1$ and and for the $I=0$ $^3P_1$, $^3P_2$ partial waves.
Here $R_1=0.9$~fm and $R_2=1.0$~fm.
The black/solid line is at N$^3$LO, the blue/dashed line at N$^2$LO,
the magenta/dotted line at NLO, and the green/dash-double-dotted line corresponds to the LO result.
\label{fig:ctg}}
\end{figure}

%
For a more quantitative assessment of the residual cutoff dependence of the
phase shifts and inelasticity parameters in a given channel we follow the
procedure described in Refs.~\cite{EKM:2015,Griesshammer:2015}. In these
works the quantity $| 1 - \cot \delta^{(R_1)}(k) / \cot \delta^{(R_2)} (k) |$
is considered as function of the cms momentum $k$, where $R_1$ and $R_2$
are two different values of the cutoff.
Since in the $\bar NN$ case the phase shifts are complex,
we examine that quantity for the real part of $\delta$
($\delta_R$) and for the imaginary part ($\delta_I$) separately,
i.e. we evaluate
$| 1 - \cot \delta^{(R_1)}_R(k) / \cot \delta^{(R_2)}_R (k) |$
and
$| 1 - \cot \delta^{(R_1)}_I(k) / \cot \delta^{(R_2)}_I (k) |$.
Corresponding results for selected partial waves can be found in Fig.~\ref{fig:ctg}
for the particular choice of $R_1=0.9\,$fm and $R_2=1.0\,$fm.

According to Ref.~\cite{EKM:2015} the residual cutoff dependence
can be viewed as an estimation of effects of higher-order contact
interactions beyond the truncation level of the potential.
Given that there are no new contact terms when going from
the chiral orders NLO and N$^2$LO, cf. Sect.~2.2, one expects
that the residual cutoff dependence reduces only when going
from LO to NLO and then again from N$^2$LO to N$^3$LO.
Indeed, the results presented in Fig.~\ref{fig:ctg} demonstrate
that the cutoff dependence at NLO and N$^2$LO is comparable.
Furthermore, there is a noticeable reduction of the cutoff dependence
over a larger momentum range when going from LO to NLO/N$^2$LO and
(in case of the $P$-waves) from NLO/N$^2$LO to N$^3$LO.
Thus, despite certain limitations, overall the behavior we observe here for
the $\bar NN$ phase shifts is similar to that in the $NN$ case~\cite{EKM:2015}.
This applies roughly also to the breakdown scale $\Lambda_b$ at N$^3$LO,
that is to the momentum at which the N$^3$LO curves cross the ones of
lower orders. In the $NN$ section it was argued that $\Lambda_b$
is about $\sim 500\,$MeV for $S$-waves and even higher for
$P$-waves~\cite{EKM:2015}. Based on the results in Fig.~\ref{fig:ctg}
we would draw a similar conclusion for the $\bar NN$ interaction.

In any case, we want to emphasize that caution has to be exercised in
the interpretation of the error plots. Specifically, one should not
forget that they provide only a qualitative guideline \cite{EKM:2015}.
In this context we want to comment also on the dips or other sharp structures
in the error plots. Those appear at
values of $k$ where the function $1 - \cot \delta^{(R_1)}(k) / \cot \delta^{(R_2)} (k) $
changes its sign or where one of the phase shifts crosses $0$ or $90$~degrees.
As already pointed out in Ref.~\cite{EKM:2015} those have no significance
and should be ignored. Indeed, a notable number of $\bar NN$ phase shifts
exhibit a strong energy dependence and, thus, cross $0$ or $90$~degrees,
cf. Figs.~\ref{fig:ph-1}-\ref{fig:ph-3}.
Because of that the kind of artefacts mentioned above occur more often in $\bar NN$,
especially in $S$-waves. Accordingly, those distort the error plots more than what
happened for the $NN$ phase shifts and make their interpretation more delicate.

\begin{figure}[htbp]
\vspace{-2.0cm}
\centering
\includegraphics[width=1.0\textwidth,height=1.0\textheight]{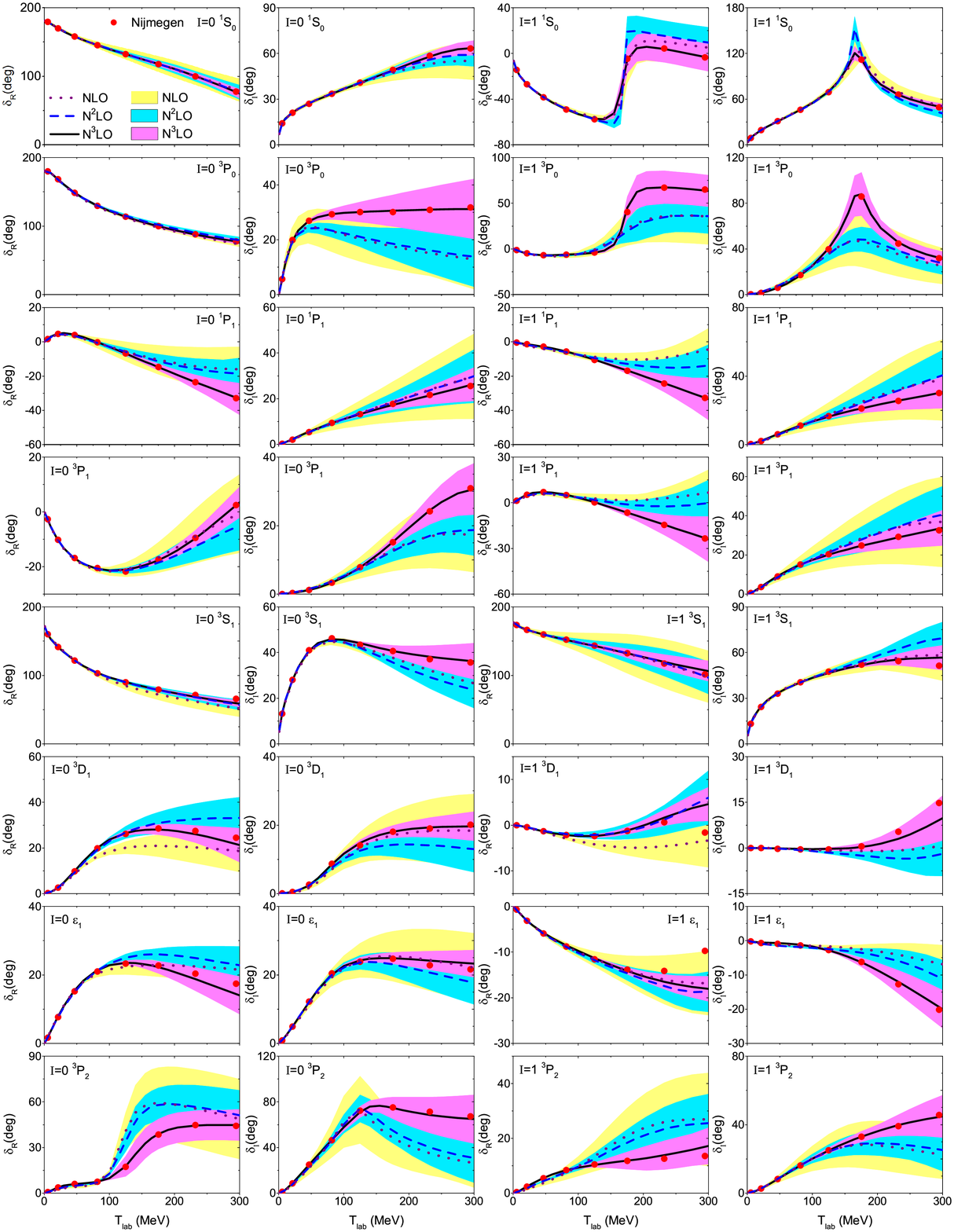}
\caption{Real and imaginary parts of various $\bar NN$ phase shifts for the potential
with cutoff $R=0.9$~fm.
Results at N$^3$LO (black/solid line), N$^2$LO (blue/dashed line), and NLO (magenta/dotted
line) are shown. Uncertainty bands
at N$^3$LO (dark/magenta), N$^2$LO (medium/cyan), and NLO (light/yellow) are included.
The filled circles represent the solution of the $\bar pp$ PWA~\cite{Zhou:2012}.
}
\label{fig:ph-band1}
\end{figure}

\begin{figure}[htbp]
\vspace{-2.0cm}
\centering
\includegraphics[width=1.0\textwidth,height=1.0\textheight]{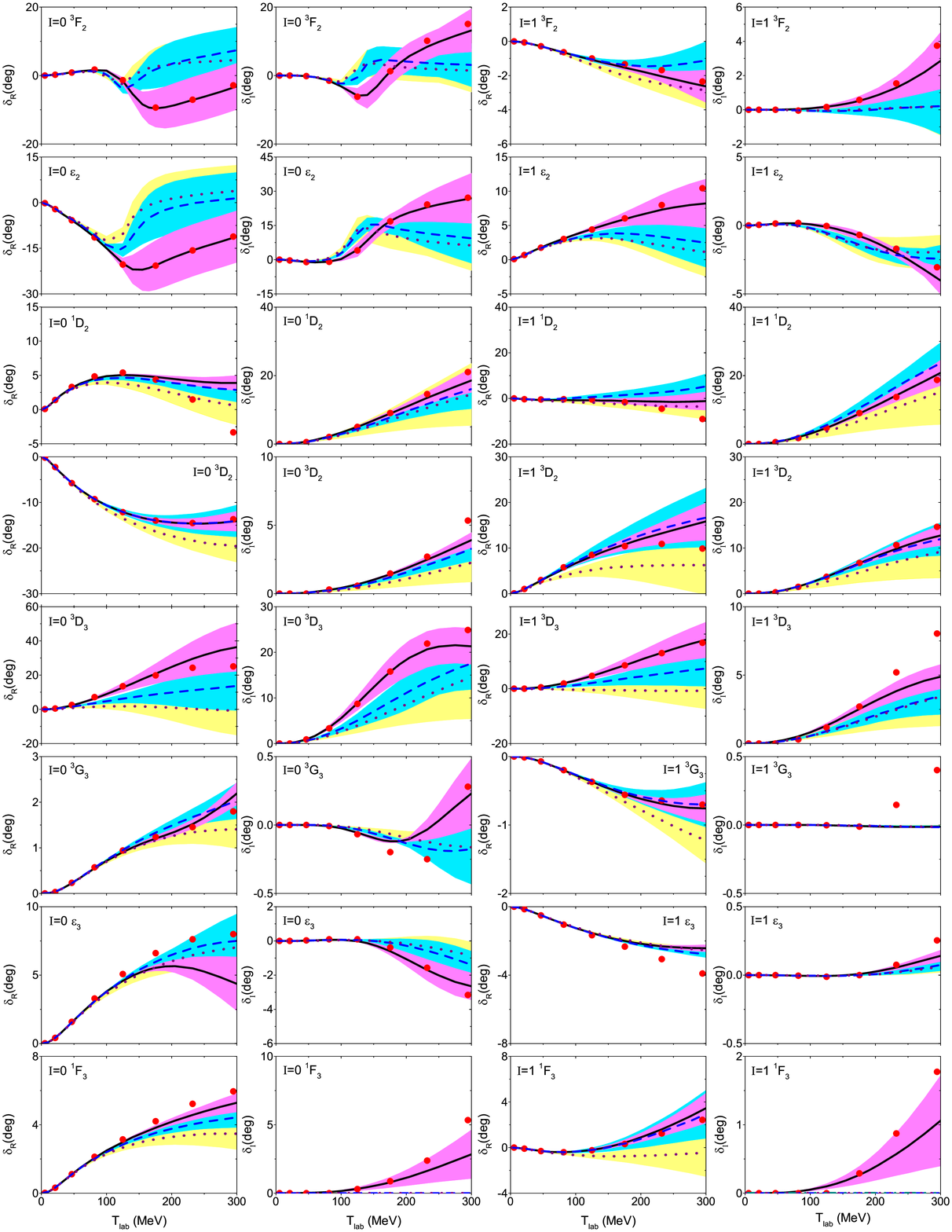}
\caption{Real and imaginary parts of various $\bar NN$ phase shifts for the potential
with cutoff $R=0.9$~fm. For notations, see Fig.~\ref{fig:ph-band1}.
}
\label{fig:ph-band2}
\end{figure}

\begin{figure}[htbp]
\vspace{-2.0cm}
\centering
\includegraphics[width=1.0\textwidth,height=0.5\textheight]{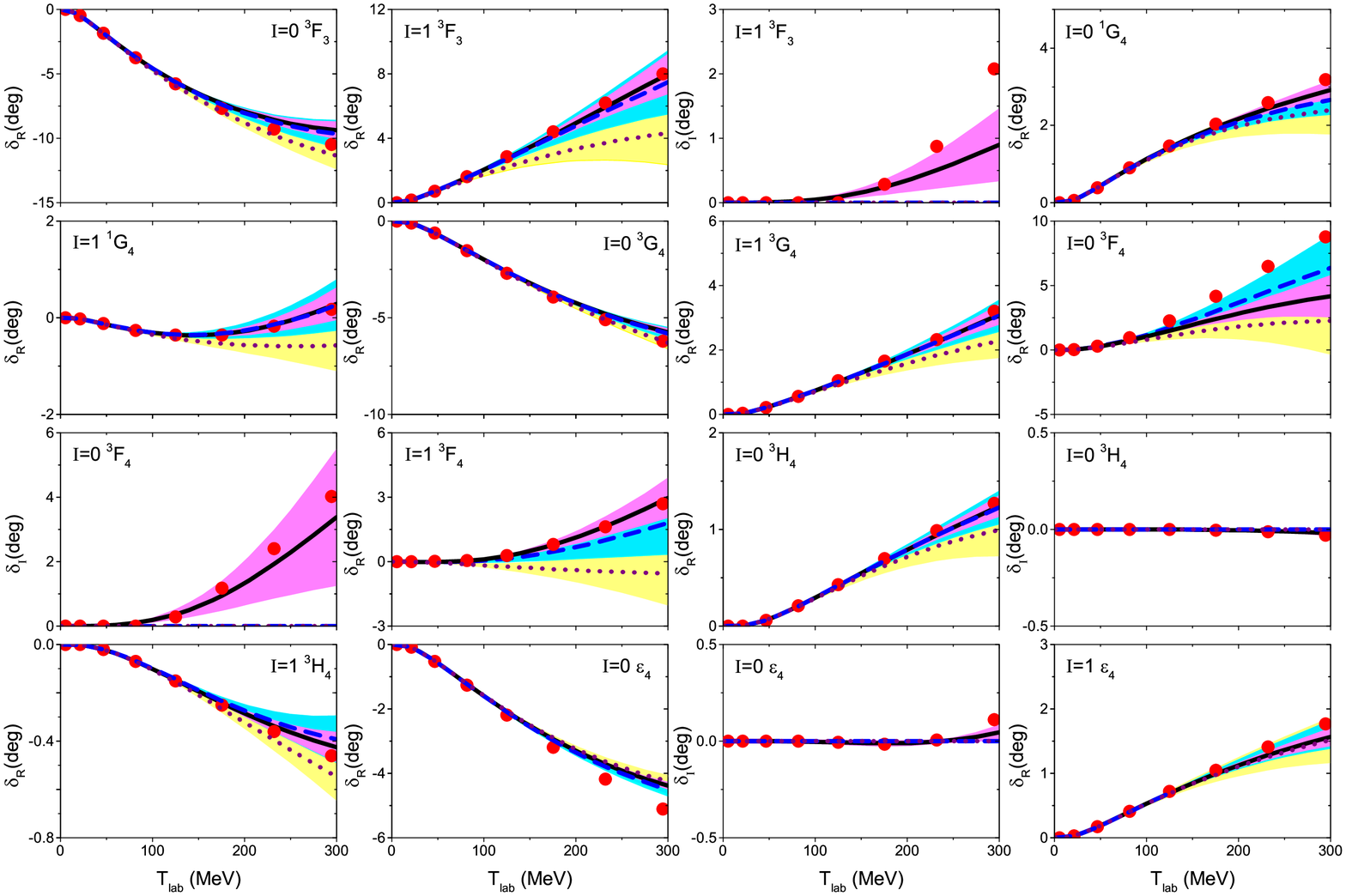}
\caption{Real and imaginary parts of various $\bar NN$ phase shifts for the potential
with cutoff $R=0.9$~fm. For notations, see Fig.~\ref{fig:ph-band1}.
}
\label{fig:ph-band3}
\end{figure}

%
The phase shifts and mixing angles for the cutoff $R=0.9\,$fm are again
presented in Fig.~\ref{fig:ph-band1}-\ref{fig:ph-band3}. However, now
results at NLO (dotted curves), N$^2$LO (dashed curves)
and N$^3$LO (solid curves) are shown and, in addition,
the uncertainty estimated via Eq.~(\ref{Error}) is indicated by bands.
The results of the $\bar NN$ PWA \cite{Zhou:2012} are displayed by circles.
There is a clear convergence visible from the curves in those figures for
most partial waves. Moreover, in case of $S$- and $P$-waves the N$^3$LO
results are in excellent agreement with the PWA in the whole considered
energy range, i.e. up to $T_{lab}=300$~MeV. This is particularly remarkable
for channels where there is a resonance-like behavior like in the
isospin $I=1$ $^1S_0$ and $^3P_0$ states, see Fig.~\ref{fig:ph-band1}.
Note that even for higher partial waves
the phase shifts and inelasticities are well described at least up to
energies of $200$ to $250$~MeV at the highest order considered, as can
be seen in Figs.~\ref{fig:ph-band2} and \ref{fig:ph-band3}.

Overall, the convergence pattern is qualitatively similar to the one for
the corresponding $NN$ partial waves reported in Ref.~\cite{EKM:2015}.
Exceptions occur, of course, in those $\bar NN$ waves where the PWA
predicts a resonance-like behavior.
Furthermore, also with regard to the uncertainty estimate, represented
by bands in Figs.~\ref{fig:ph-band1}-\ref{fig:ph-band3}, in general,
the behavior resembles the one observed in the application of
chiral EFT to $NN$ scattering.
Specifically, it is reassuring to see that in most cases
also for $\bar NN$ the uncertainty as defined in Eq.~(\ref{Error})
fulfills the conditions and expectations discussed in Sect.~7 of
Ref.~\cite{EKM:2015}.
Thus, we conclude that the approach for error estimation suggested in
Ref.~\cite{EKM:2015} is well applicable for the $\bar NN$ case, too.

Some more detailed observations: It is interesting to see that
in the $^1S_0$, $^3P_0$ and $^3S_1$ partial waves with $I=0$
the uncertainty is very small, even at $T_{lab}=300$~MeV, just like what
was found for the corresponding $NN$ states.
On the other hand, and not unexpected, there is a much larger uncertainty
in the $I=1$ state, in particular in the $^1S_0$ and $^3P_0$ waves.
Again this has to do with the resonance-like behavior.
As noted above, these structures can be reproduced quantitatively only
at the highest order and the poorer convergence in this case is then
reflected in a larger uncertainty - as it should be according to its
definition, see Eq.~(\ref{Error}).
Such a resonance-like behavior and/or an ``unusually'' strong energy
dependence at higher energies of phase shifts is also the main reason
why for some cases the uncertainty estimate fails to produce the desired
results, i.e. where the bands do not show a monotonic behavior, where they
do not overlap for different orders, or where the PWA results lie outside of
the uncertainty bands. Examples for that are the inelasticity for
$^1S_0$ with $I=1$, the inelasticity for $^3P_0$ with $I=1$, or the
$^3P_2$ and $^3F_2$ phase shifts and the mixing angle $\epsilon_2$ with $I=0$.
Note that in many cases there is a larger uncertainy for the inelasticity
than for the phase shift itself. Again this is not unexpected. For
$P$- and higher partial waves nonzero results for the inelasticity
are only obtained from NLO onwards in the power counting we follow
so that the convergence is slower.
Finally, let us mention that in some $F$-, $G$-, and $H$-waves the
inelasticity is zero or almost zero \cite{Zhou:2012}. We omitted the
corresponding graphs from Fig.~\ref{fig:ph-band3}.

\subsection{Observables}

In our first study of $\bar NN$ scattering within chiral EFT \cite{Kang:2013}
we focused on the phase shifts and inelasticities. Observables were not
considered. One reason for this was that, at that time, our computrt code was
only suitable for calculations in the isospin basis. A sensible calculation
of observables, specifically at low energies where chiral EFT should work
best, has to be done
in the particle basis because the Coulomb interaction in the $\bar pp$
system has to be taken into account and also the mass difference between
proton and neutron. The latter leads to different physical thresholds for
the $\bar pp$ and $\bar nn$ channels which has a strong impact on the
reaction amplitude close to those thresholds.

Another reason is related directly to the dynamics of $\bar NN$ scattering,
specifically to the presence of annihilation processes.
Annihilation occurs predominantly at short distances and yields a reduction
of the magnitude of the $S$-wave amplitudes. Because of that, higher partial
waves start to become important at much lower energies as compared to what one
knows from the $NN$ interaction \cite{Rev3}.
Thus, already at rather moderate energies a realistic description of higher
partial waves, in particular of the $P$- as well as $D$-waves, is required
for a meaningful confrontation of the computed amplitudes with scattering data.

\begin{figure}[htbp]
\centering
\includegraphics[width=0.9\textwidth,height=0.50\textheight]{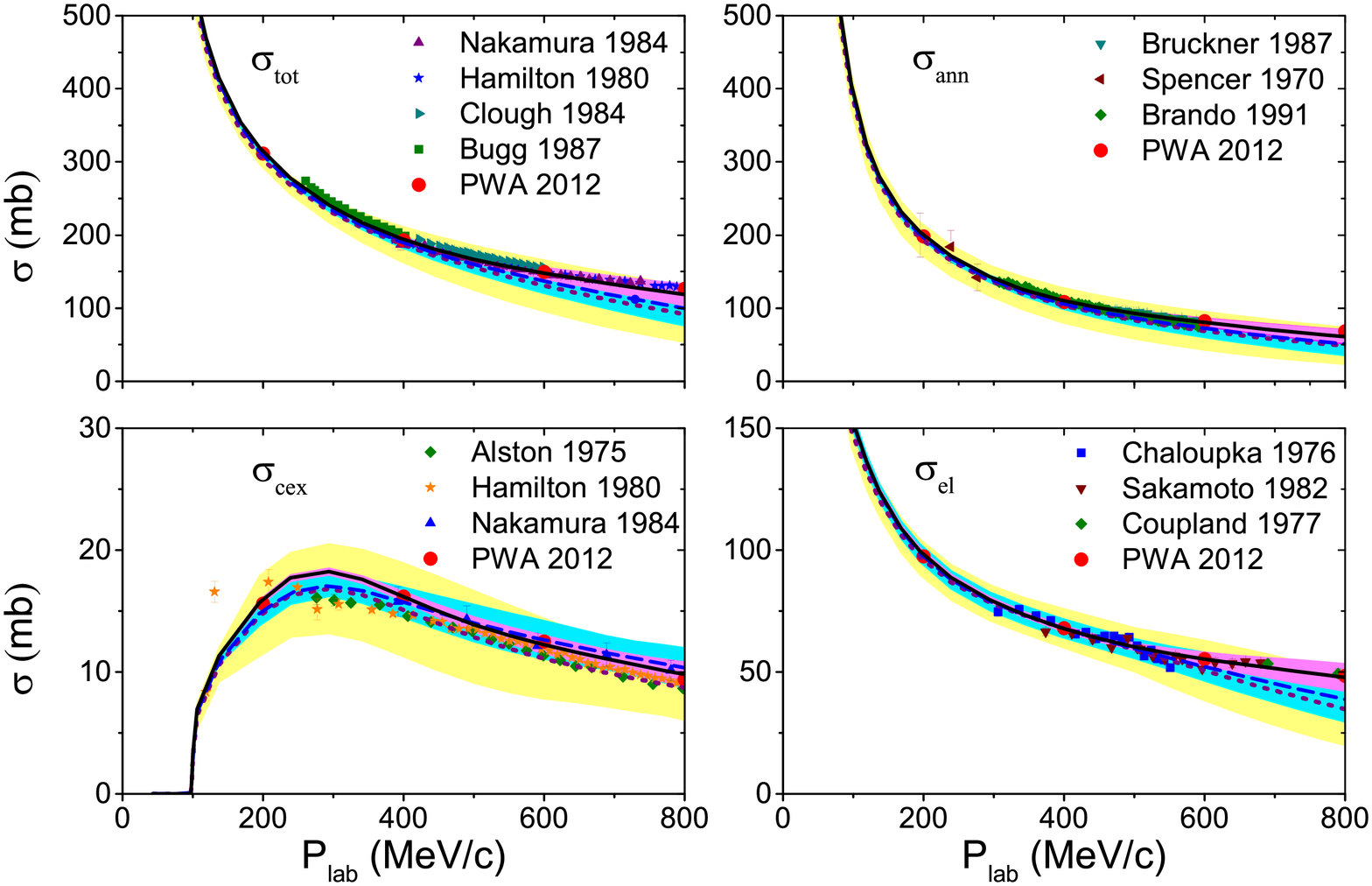}
\caption{Total ($\sigma_{tot}$) and integrated elastic ($\sigma_{el}$), charge-exchange
($\sigma_{cex}$), and annihilation ($\sigma_{ann}$) cross sections
for $\bar pp$ scattering.
Results at N$^3$LO (black/solid line), N$^2$LO (blue/dashed line), and NLO (magenta/dotted
line) are shown. Uncertainty bands at N$^3$LO (dark/magenta), N$^2$LO (medium/cyan),
and NLO (light/yellow) are included.
The filled circles represent the solution of the $\bar pp$ PWA~\cite{Zhou:2012}.
Data are taken from Refs.
\cite{Nakamura1984prd,Hamilton1980prl,Clough1984plb,Bugg1987plb} ($\sigma_{tot}$),
\cite{Bruckner1987plb,Spencer1970,Brando1985}
($\sigma_{ann}$),
\cite{Alston1975,Hamilton1980prl2,Nakaruma1984prl}
($\sigma_{cex}$), and
\cite{Chaloupka1976,Sakamoto1982,Coupland1977}
($\sigma_{el}$).
}
\label{fig:cs}
\end{figure}

In the present paper we extended our chiral EFT $\bar NN$ potential to N$^3$LO.
At that order the first LECs in the $D$-waves appear, cf. Eq.~(\ref{VC}),
and can be used to improve
substantially the reproduction of the corresponding partial-wave amplitudes of
the $\bar NN$ PWA, cf. Figs.~\ref{fig:ph-band1} and \ref{fig:ph-band2}.
Thus, it is now timely to perform also a calculation of observables and compare
those directly with measurements. Integrated cross sections are shown in
Fig.~\ref{fig:cs}. Results are provided for the total reaction cross
section, for the total annihilation cross section, and for the integrated
elastic ($\bar pp \to \bar pp$) and charge-exchange ($\bar pp \to \bar nn$) cross sections.
Similar to the presentation of the phase shifts before, we include curves
for the NLO (dotted lines), N$^2$LO (dashed lines), and N$^3$LO (solid lines)
results and indicate the corresponding uncertainty
estimate by bands for the cutoff $R=0.9\,$fm.
The LO calculation is not shown because it provides only a very limited
and not realistic description of observables. Instead we include a variety of
experimental results.

Before discussing the results in detail let us make a general comment on the data.
We display experimental information primarily for illustrating the overall quality of
our results. Thus, we choose specific measurements at specific energies which fit
best to that purpose, and we use the values as published in the original papers.
This differs from the procedure in the PWA \cite{Zhou:2012} where data selection
is done and has to be done. After all, one cannot do a dedicated PWA without having
a self-consistent data set. Thus, normalization factors are introduced for the data
sets in the course of the PWA and some data have been even rejected. For details
on the criteria employed in the PWA and also for individual
information on which data sets have been renormalized or rejected we refer the reader
to Ref.~\cite{Zhou:2012}.
In view of this it is important to realize that there can be cases where our EFT
interaction reproduces the PWA perfectly but differs slightly from the real data (when a
renormalization was employed) or even drastically (when those data were rejected).
Of course, in the latter case we will emphasize that in the discussion.

Our results for the integrated cross sections at N$^3$LO, indicated by solid
lines in Fig.~\ref{fig:cs},
agree rather well with the ones of the PWA (filled circles), even up to $p_{lab}=800$~MeV/c.
Indeed, also the charge-exchange cross section is nicely reproduced, though it is much smaller
than the other ones. The amplitude for this process is given by the difference of
the $I=0$ and $I=1$ amplitudes and its description requires a delicate balance between
the interactions in the corresponding isospin channels.
Obviously, this has been achieved with our chiral EFT interaction.
Note that there are inconsistencies in the charge-exchange measurements at low energies
and some of the data in question have not been taken into account in the PWA, cf. Table~III
in \cite{Zhou:2012}.
Considering the bands presenting the estimate of the uncertainty, one can see that there is a
clear convergence of our results for all cross sections when going to higher orders.
Finally, as a further demonstration of the quality of our N$^3$LO results we summarize
partial-wave cross sections for $\bar pp$ elastic and charge-exchange scatting in
Table~\ref{cs:par}.
Obviously, there is nice agreement with the values from the PWA for basically all
$S$- and $P$-waves.

\begin{table}[h]
\renewcommand{\arraystretch}{1.10}
\centering
\caption{Partial-wave cross sections predicted by the chiral potential at N$^3$LO with $R=0.9$~fm
in comparison to results from the $\bar NN$ partial wave analysis \cite{Zhou:2012}.}
\label{cs:par}
\begin{tabular}{|c|c||rrrr||rrrr|}
\hline
&& \multicolumn{4}{|c||}{$\bar p p \to \bar p p$} & \multicolumn{4}{|c|}{$\bar p p \to \bar n n$} \\
& $p_{lab}$ (MeV/c) & 200 & 400 & 600& 800 & 200 & 400 & 600& 800 \\
\hline
\hline
\multirow{2}{*}{$^1S_0$}
        & {N$^3$LO}      & 15.9 & 8.0 & 4.1 & 2.0 & 0.7 & 0.1 &     &  \\
        & {PWA}          & 15.7 & 7.9 & 4.1 & 2.1 & 0.7 & 0.1 &     &  \\
\hline
\multirow{2}{*}{$^3S_1$}
        & {N$^3$LO}      & 66.6 &25.9 &13.1 & 8.0 & 2.9 & 0.9 & 0.5 & 0.3 \\
        & {PWA}          & 66.1 &26.0 &13.2 & 8.8 & 3.0 & 1.0 & 0.5 & 0.2 \\
\hline
\multirow{2}{*}{$^3P_0$}
        & {N$^3$LO}      &  4.9 & 5.4 & 5.1 & 3.6 & 1.5 & 0.8 & 0.1 &     \\
        & {PWA}          &  4.9 & 5.4 & 5.0 & 3.5 & 1.5 & 0.8 & 0.1 &     \\
\hline
\multirow{2}{*}{$^1P_1$}
        & {N$^3$LO}      &  1.0 & 2.5 & 4.4 & 5.6 & 0.8 & 0.1 &     &     \\
        & {PWA}          &  0.9 & 2.5 & 4.5 & 5.6 & 0.8 & 0.1 &     &     \\
\hline
\multirow{2}{*}{$^3P_1$}
        & {N$^3$LO}      &  1.8 & 5.0 & 4.1 & 3.6 & 5.1 & 3.0 & 0.2 & 0.1 \\
        & {PWA}          &  1.8 & 4.9 & 4.0 & 3.5 & 4.9 & 2.9 & 0.2 & 0.1 \\
\hline
\multirow{2}{*}{$^3P_2$}
        & {N$^3$LO}      &  7.0 &17.1 &14.1 &9.9  & 1.0 & 1.5 & 0.4 & 0.1 \\
        & {PWA}          &  7.0 &17.0 &13.9 &9.6  & 0.9 & 1.4 & 0.4 & 0.1 \\
\hline
\end{tabular}
\end{table}

\begin{figure}[htbp]
\vspace{-2.0cm}
\centering
\includegraphics[width=1.0\textwidth,height=1.0\textheight]{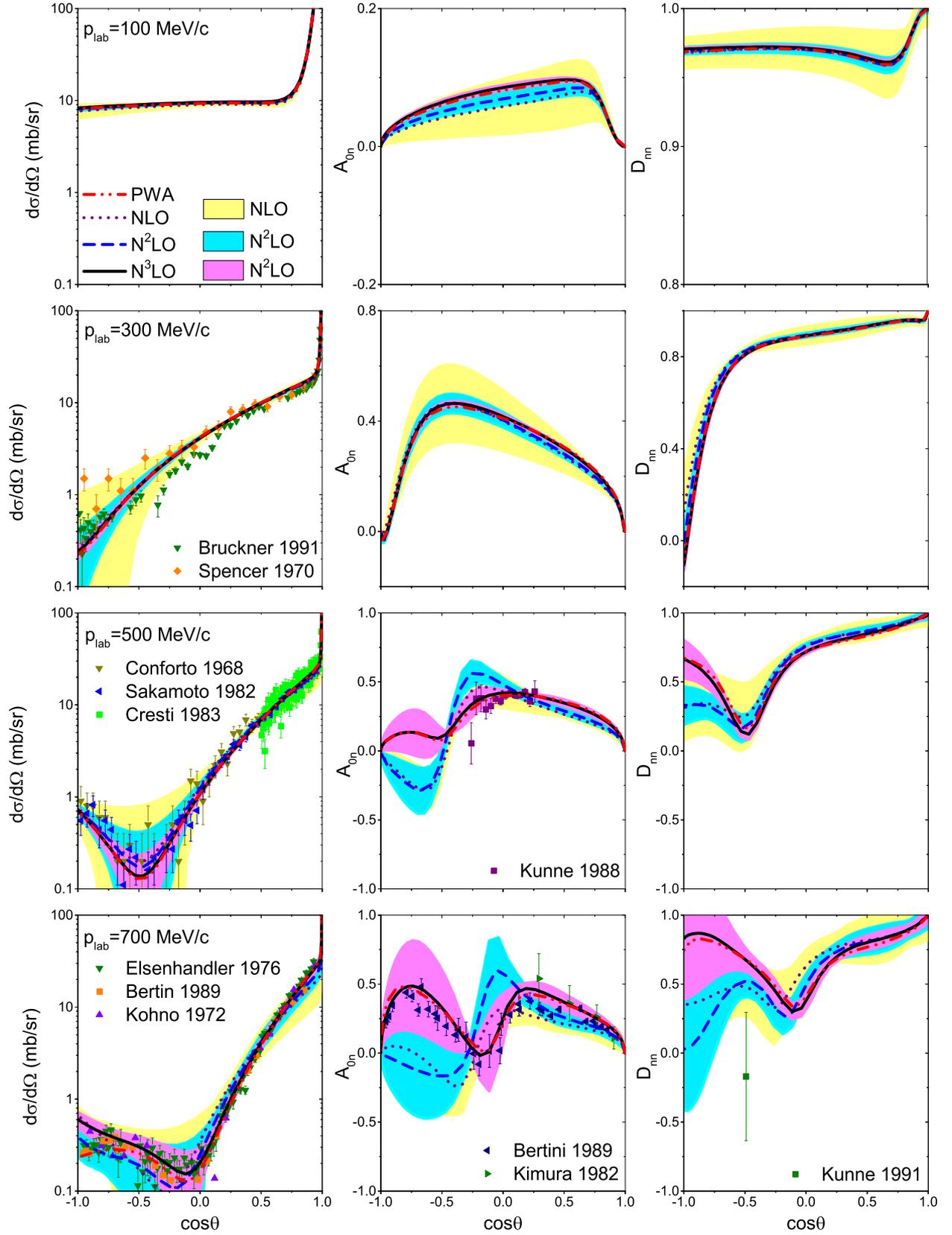}
\caption{Differential cross sections, analyzing powers and spin correlation parameters $D_{nn}$
for $\bar pp$ elastic scattering.
For notations, see Fig.~\ref{fig:cs}.
The red/dash-double dotted line represents the result of the PWA~\cite{Zhou:2012}.
Data are taken from Refs.
\cite{Bruckner1991,Spencer1970,Conforto1968,Sakamoto1982,Cresti1983,Eisenhandler1976,Bertin1989,Kohno1972}
(differential cross sections),
\cite{Kunne1988,Bertini1989,Kimura1982} (analyzing powers),
and \cite{Kunne1991} ($D_{nn}$).
}
\label{fig:dcs-cel}
\end{figure}

\begin{figure}[htbp]
\vspace{-2.0cm}
\centering
\includegraphics[width=1.0\textwidth,height=0.5\textheight]{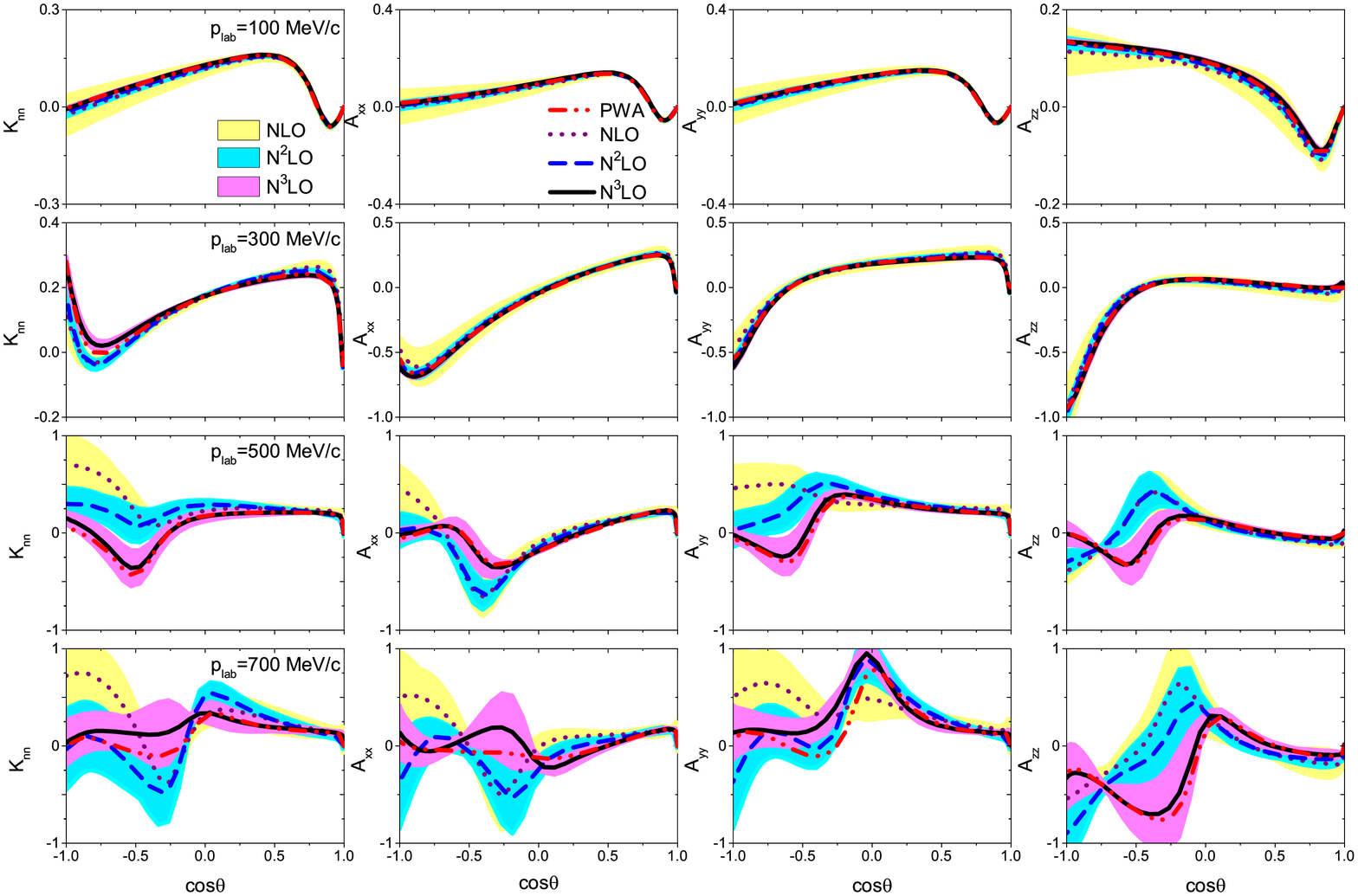}
\caption{Spin correlation parameters $K_{nn}$, $A_{xx}$, $A_{yy}$, and $A_{zz}$
for $\bar pp$ elastic scattering.
For notations, see Fig.~\ref{fig:dcs-cel}.
}
\label{fig:Knn-cel}
\end{figure}

Differential cross sections, analyzing powers and the spin-correlation parameters $D_{nn}$
for $\bar pp$ elastic scattering are shown in Fig.~\ref{fig:dcs-cel}.
Results for further spin-dependent observables can be found in Fig.~\ref{fig:Knn-cel}.
We selected results at the momenta $100$, $300$, $500$, and $700$~MeV/c
($T_{lab}=$~$5.32$, $46.8$, $125$, and $232$~MeV) for the presentation
because that allows us to compare with some existing measurements (for $d\sigma/d\Omega$, $A_{on}$)
and it allows us also to document how the quality of the description of $\bar NN$
scattering observables by our EFT interaction develops with increasing energy.
The results of the $\bar NN$ PWA \cite{Zhou:2012} are indicated by dash-dotted lines.
Since only $\bar NN$ partial waves up to $J=4$ are tabulated in Ref.~\cite{Zhou:2012}
we supplemented those by amplitudes from our N$^3$LO interaction for higher angular momenta
in the evaluation of differential observables. As already emphasized above, those amplitudes
differ to some extent from the ones used in the PWA itself. But we do not expect that
those differences have a strong influence on the actual results.
Note that contributions from $J\ge 5$ become relevant for momenta above $400$~MeV/c, but
primarily at backward angles.

In principle, at the lowest energy considered, $T_{lab}=5.32$~MeV, we expect excellent
agreement of our calculation with the PWA. However, one has to keep in mind that we
fitted to the phase shifts and inelasticies in the isospin basis. The observables
are calculated from partial-wave amplitudes in the particle basis. The
latter are obtained by solving the corresponding LS equation where then
the hadronic interaction is modified due to the presence of the Coulomb interaction,
and there are additional kinematical effects  from the shift of the $\bar nn$
threshold to its physical value.
Therefore, it is not trivial that we agree so well with the PWA results, that are generated
from the $S$-matrix elements in the particle basis as listed in Ref.~\cite{Zhou:2012}.
Actually, in case of the differential cross section
one cannot distinguish the corresponding (solid, dash-dotted) lines in the figure.
The estimated uncertainty is also rather small at least for the differential
cross section. Spin-dependent observables involve contributions from higher partial
waves from the very beginning and because of that the uncertainties are larger,
especially for the lower-order results. There is no experimental information on
differential observables at such low energies.

Naturally, when we go to higher energies the uncertainty increases. In this context
we want to point out that the differential cross section exhibits a rather strong
angular dependence already at $p_{lab}=300$~MeV/c. Its value drops by more than one
order of magnitude with increasing angles, cf. Fig.~\ref{fig:dcs-cel}.
This means that at backward angles there must be a delicate cancellation between
many partial-wave amplitudes and, accordingly, a strong sensitivity to the accuracy
achieved in each individual partial wave. Note also that a logarithmic scale
is used that optically magnifies the size of the uncertainty bands for small
values.
The behavior of $d\sigma/d\Omega$ for the $\bar pp$ reaction differs considerably
from the one for $NN$ scattering where the angular dependence is relatively weak,
even at higher energies~\cite{EKM:2015}. In fact the features seen in $\bar pp$ scattering
are more comparable with the ones for nucleon-deuteron ($Nd$) scattering, see
e.g. the results in Ref.~\cite{Binder:2015}.

\begin{figure}[htbp]
\vspace{-2.0cm}
\centering
\includegraphics[width=1.0\textwidth,height=1.0\textheight]{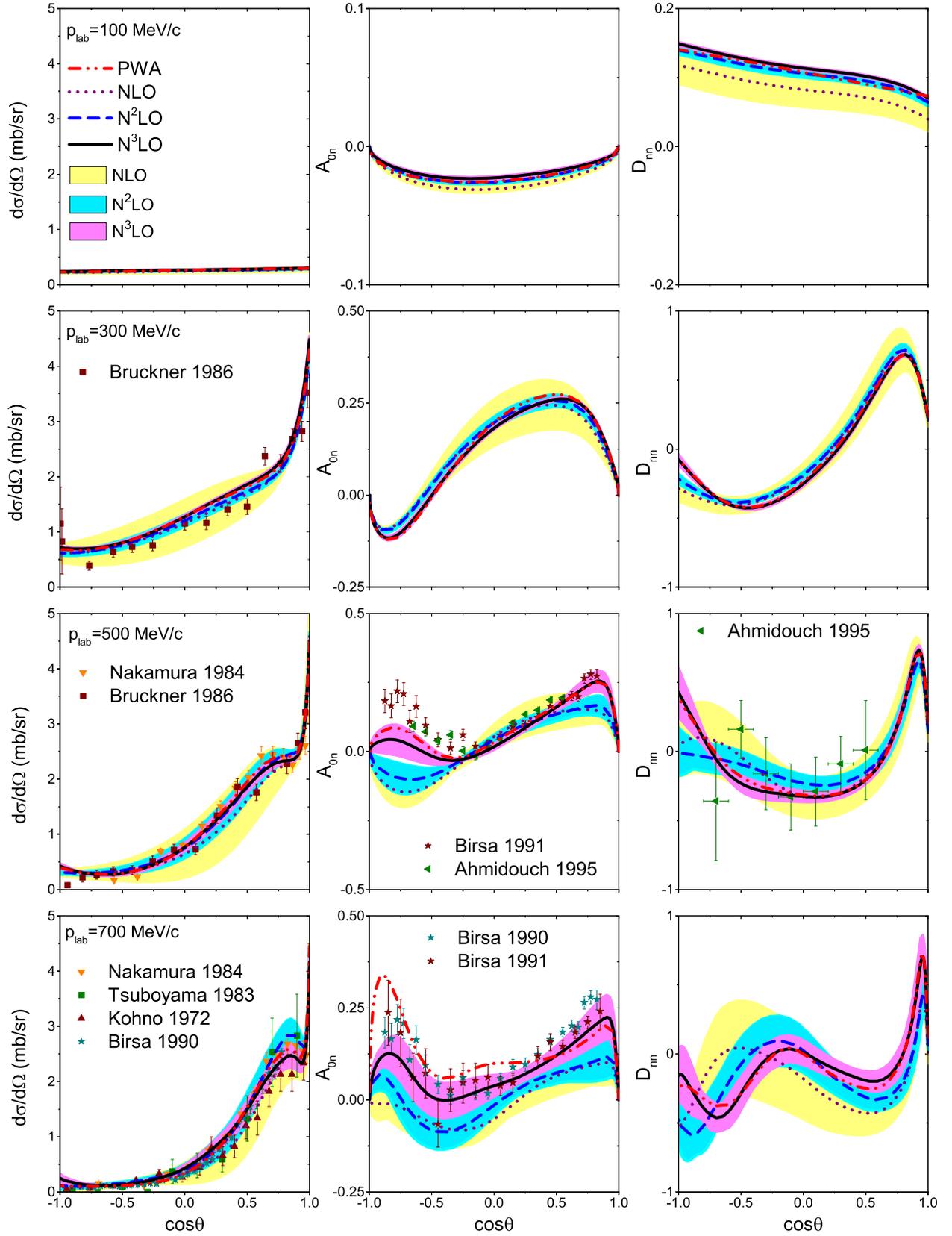}
\caption{Differential cross sections, analyzing powers and spin correlation parameters $D_{nn}$
for charge-exchange scattering.
For notations, see Fig.~\ref{fig:dcs-cel}.
Data are taken from Refs.
\cite{Bruckner1986plb2,Nakaruma1984prl,Tsuboyama1983,Kohno1972,Birsa1990}
(differential cross sections),
\cite{Birsa1991,Ahmidouch1995,Birsa1990}.
(analyzing powers), and \cite{Ahmidouch1995} ($D_{nn}$).
Note that the data for $A_{on}$ are for $546$ and $656$ MeV/c, respectively.
}
\label{fig:dcs-cex}
\end{figure}

\begin{figure}[htbp]
\vspace{-2.0cm}
\centering
\includegraphics[width=1.0\textwidth,height=0.5\textheight]{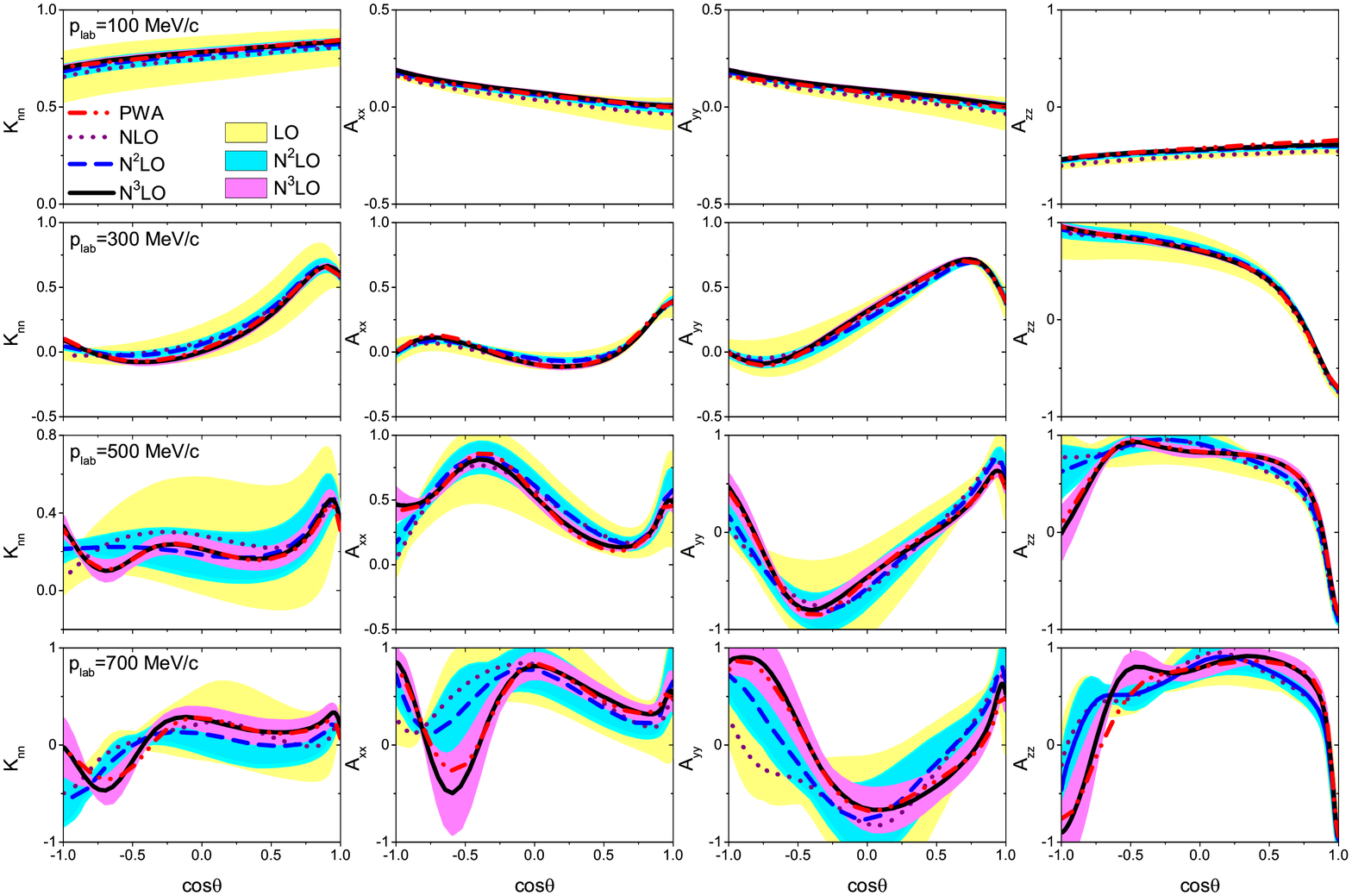}
\caption{Spin correlation parameters $K_{nn}$, $A_{xx}$, $A_{yy}$, and $A_{zz}$
for charge-exchange scattering.
For notations, see Fig.~\ref{fig:dcs-cel}.
}
\label{fig:Knn-cex}
\end{figure}

Also with regard to the analyzing power $A_{on}$ the uncertainty bands look similar
to the pattern one observes in $Nd$ scattering. As already said above, for spin-dependent
observables higher partial waves play a more important role and the uncertainty in
their reproduction is also reflected more prominently in the results for the
observables. Interestingly, the uncertainty exhibits a strong angular dependence.
It seems that the angles where it is small are strongly correlated with the zeros
of specific Legendre polynomials where then the contributions of, say, $D$-waves
are zero and likewise their contribution to the uncertainty. For $A_{on}$ and also for
other spin-dependent observables there is a visible difference between our
N$^3$LO results (solid curve) and the PWA (dash-dotted curve) at the highest
energy displayed in Figs.~\ref{fig:dcs-cel} and \ref{fig:Knn-cel}.

Differential cross sections, analyzing powers and the spin-correlation parameters $D_{nn}$
for the charge-exchange reaction $\bar pp \to \bar nn$ are shown in Fig.~\ref{fig:dcs-cex}.
Results for further spin-dependent observables can be found in Fig.~\ref{fig:Knn-cex}.
The quality of the reproduction of the PWA results by our EFT interaction at
N$^3$LO but also the convergence properties with increasing order and the
uncertainties are similar to those observed for $\bar pp$ elastic scattering.
However, visible deviations from the PWA start already at somewhat smaller energies.
This is most obvious in case of the analyzing power $A_{on}$ where noticeable differences of
our N$^3$LO results to those of the PWA occur already from
$p_{\rm lab} \sim 500\,$MeV ($T_{\rm lab} \sim 125\,$MeV) onwards, cf.
Fig.~\ref{fig:dcs-cex}. Note that the lowest momentum is very close to the
$\bar nn$ threshold, which is at $p_{\rm lab} = 98.70\,$MeV, so that the
kinetic energy in the $\bar nn$ system is only of the order of a few keV.
Despite of that the spin-dependent observables exhibit already a distinct angular
dependence and $A_{on}$ is clearly nonzero.

In any case, overall we can conclude that chiral EFT at N$^3$LO not only allows for
an excellent reproduction of the PWA results but also of the actual observables
for energies below $p_{\rm lab} \sim 500\,$MeV ($T_{\rm lab} \sim 125\,$MeV)
and it still provides a good description of the data at energies of the order of
$p_{\rm lab} \sim 700\,$MeV ($T_{\rm lab} \sim 230\,$MeV)

\section{Predictions}

The lowest momentum for which results of the PWA are provided in Ref.~\cite{Zhou:2012},
and accordingly are taken into account in our fitting procedure,
is $p_{lab} = 100$~MeV/c corresponding to $T_{lab} = 5.32$~MeV.
As can be seen in Table~III of Ref.~\cite{Zhou:2012} no data below $100$~MeV/c
have been included in the analysis, and only a few below $200$~MeV/c.
In view of this we consider results of our $\bar NN$ potential at momenta below
$100$~MeV/c as genuine predictions. First of all this concerns the low-energy
structure of the amplitudes given in terms of the effective range
expansion.  Results for the scattering lengths (for $^1S_0$ and $^3S_1$) and
for scattering volumes (for the $P$ waves) are summarized in
Table~\ref{tab:a_iso}. These are complex numbers because of the presence
of annihilation.
The pertinent calculations were done in the isospin basis and the isospin $I$
is included here in the spectral notation, i.e. we write $^{(2I+1)(2S+1)}L_J$.
As one can see in Table~\ref{tab:a_iso} the results for the $^1S_0$ partial waves
are very stable and change very little with increasing order. There is a slightly
larger variation in case of the $^3S_1$. Somewhat stronger variations occur
in the $P$ waves where those in the $^3P_2$ partial waves are by far the most
dramatic ones. This is not surprising in view of the coupling of the $^3P_2$ to
the $^3F_2$ and the fact that there is only a single (but complex-valued) LEC at NLO and
N$^2$LO that can be used in the fit to the $^3P_2$ and $^3F_2$ phase shifts and
the mixing angle $\epsilon_2$.

\begin{table}[htbp]
\begin{center}
\caption{Scattering lengths and volumes for different partial waves for the chiral potentials
with $R=0.9$~fm. $\bar{a}^{cs}_{S,\bar{p}p}$ and $\bar{a}^{cs}_{P,\bar{p}p}$ are
spin-averaged results obtained from a calculation in particle basis including the
Coulomb force.
For comparison N$^2$LO predictions of our previous chiral potential are included, based on the
cutoff combination ($\Lambda$, $\tilde\Lambda$) = (450,500)~MeV \cite{Kang:2013}.
}
\label{tab:a_iso}
 \vspace{0.3cm}
\renewcommand{\arraystretch}{1.3}
\begin{tabular}{|c||c|c|c|c|}
\hline\hline
\rule[0.5cm]{0.2cm}{0cm}\rule[0.5cm]{0.2cm}{0cm} & \rule{0.28cm}{0cm} NLO \rule{0.28cm}{0cm}    & \rule{0.28cm}{0cm} N$^2$LO\rule{0.28cm}{0cm}     &\rule{0.28cm}{0cm} N$^3$LO\rule{0.28cm}{0cm}  &   N$^2$LO \cite{Kang:2013} \\
\hline\hline
$a_{^{11}S_0}$~(fm)       &  $-$0.21~$-$~i 1.20    & $-$0.21~$-$~i 1.22     & $-$0.20~$-$~i 1.23    &$-$0.21~$-$~i1.21    \\ \hline
$a_{^{31}S_0}$~(fm)       &     1.06~$-$~i 0.57    &    1.05~$-$~i 0.60     &    1.05~$-$~i 0.58    &   1.03~$-$~i0.58    \\ \hline
$a_{^{13}S_1}$~(fm)       &     1.33~$-$~i 0.85    &    1.39~$-$~i 0.89     &    1.42~$-$~i 0.88    &   1.37~$-$~i0.88    \\ \hline
$a_{^{33}S_1}$~(fm)       &     0.44~$-$~i 0.92    &    0.45~$-$~i 0.95     &    0.44~$-$~i 0.96    &   0.44~$-$~i0.91    \\ \hline
$a_{^{13}P_0}$~(fm$^3$)   &  $-$3.62~$-$~i 8.05    & $-$3.18~$-$~i 8.02     & $-$2.83~$-$~i 7.82    &$-$3.76~$-$~i7.16    \\ \hline
$a_{^{33}P_0}$~(fm$^3$)   &     2.22~$-$~i 0.31    &    2.16~$-$~i 0.32     &    2.18~$-$~i 0.19    &   2.36~$-$~i1.14    \\ \hline
$a_{^{11}P_1}$~(fm$^3$)   &  $-$2.72~$-$~i 0.34    & $-$2.76~$-$~i 0.35     & $-$2.87~$-$~i 0.36    &$-$2.87~$-$~i0.25    \\ \hline
$a_{^{31}P_1}$~(fm$^3$)   &     0.97~$-$~i 0.29    &    0.87~$-$~i 0.31     &    0.80~$-$~i 0.34    &   0.86~$-$~i0.20    \\ \hline
$a_{^{13}P_1}$~(fm$^3$)   &     4.65~$-$~i 0.07    &    4.60~$-$~i 0.07     &    4.61~$-$~i 0.05    &   4.77~$-$~i0.02    \\ \hline
$a_{^{33}P_1}$~(fm$^3$)   &  $-$1.81~$-$~i 0.47    & $-$1.92~$-$~i 0.50     & $-$2.04~$-$~i 0.55    &$-$2.02~$-$~i0.39    \\ \hline
$a_{^{13}P_2}$~(fm$^3$)   &  $-$0.42~$-$~i 0.96    & $-$0.55~$-$~i 1.03     & $-$0.74~$-$~i 1.13    &$-$0.45~$-$~i0.57    \\ \hline
$a_{^{33}P_2}$~(fm$^3$)   &  $-$0.29~$-$~i 0.37    & $-$0.38~$-$~i 0.38     & $-$0.48~$-$~i 0.34    &$-$0.28~$-$~i0.23    \\
\hline\hline
$\bar{a}^{cs}_{S,\bar{p}p}$~(fm)  & 0.78~$-$~i~0.71   & 0.80~-~i~0.73   & 0.80~$-$~i~0.74  & 0.79~$-$~i~0.72  \\ \hline
$\bar{a}^{cs}_{P,\bar{p}p}$~(fm$^3$)  & $-$0.05~$-$~i0.74 & $-$0.12~-~i 0.77& $-$0.19~$-$~i 0.77 & $-$0.10~$-$~i0.55  \\
\hline\hline
\end{tabular}
\end{center}
\renewcommand{\arraystretch}{1.0}
\end{table}

Table~\ref{tab:a_iso} contains also scattering lengths and volumes predicted in
our earlier study
of the $\bar NN$ interaction within chiral EFT  based on a momentum-space
cutoff~\cite{Kang:2013}.
We include here the results at N$^2$LO and for the cutoff combination
($\Lambda$,$\tilde\Lambda$) = ($450$,$500$)~MeV. It is reassuring to see that in most partial
waves the predictions are very similar or even identical. More noticeable differences
occur only in $P$ waves, and in particular in the $^3P_2$ -- for the reasons just
discussed.

There is some experimental information that puts constraints on these
scattering lengths. Measurements of the level shifts and widths of
antiproton-proton atoms have been used to infer values for the spin-averaged
$\bar pp$ scattering lengths. Corresponding results can be found in Ref.~\cite{Gotta},
together with values for the imaginary part of the scattering lengths that are
deduced from measurements of the $\bar np$ annihilation cross section in
combination with the ones for $\bar pp$ annihilation.
Here we prefer to compare our predictions directly with
the measured level shifts and widths \cite{Ziegler:1988bp,Augsburger:1999yt,
Heitlinger:1991cn,Gotta:1999vj}, see Table~\ref{tab4}.
For that the Trueman formula \cite{Trueman} was applied to the theory results
with the second-order term taken into account for the $S$-waves. It has been
found in Ref.~\cite{Carbonell} that values obtained in this way agree rather
well with direct calculations. In this context let us recall
that the results in Table~\ref{tab4}, including those for the N$^2$LO interaction from
Ref.~\cite{Kang:2013}, are deduced, of course, from a calculation in particle basis.
In particular, the Coulomb force in $\bar pp$ is taken into account and likewise
the $p$-$n$ mass difference that leads to separated thresholds for the $\bar pp$
and $\bar nn$ channels. The corresponding results given in our earlier study of
the $\bar NN$ interaction within chiral EFT \cite{Kang:2013} are from a
calculation in the isospin basis.

Experimental evidence on level shifts and widths in antiprotonic hydrogen was
not taken into account in the PWA~\cite{Zhou:2012}. Anyway, it should be said
that additional assumptions have to be made in order to derive the splitting
of the $^1S_0$ and $^3S_1$ level shifts from the experiment \cite{Gotta:1999vj,GottaP}.
This caveat has to be kept in mind when comparing the theory results with
experiments. Notwithstanding, there is a remarkable agreement between our predictions
and the experimental values, with the only exception being the level shift in
the $^3P_0$ partial wave.

\begin{table}[htbp]
\begin{center}
\caption{Hadronic shifts and broadenings in hyperfine states of $\bar p$H for the chiral potentials
with $R=0.9$~fm.
For comparison N$^2$LO predictions of our previous chiral potential are included, based on the
cutoff combination ($\Lambda$, $\tilde\Lambda$) = (450,500)~MeV \cite{Kang:2013}.
The experimental information is taken from
Refs.~\cite{Ziegler:1988bp,Heitlinger:1991cn,Augsburger:1999yt,Gotta:1999vj}.
}
\label{tab4}
\vspace{0.3cm}
\renewcommand{\arraystretch}{1.2}
\begin{tabular}{|c||c|c|c|c|c|}
\hline\hline
\rule[0.5cm]{0.2cm}{0cm}\rule[0.5cm]{0.2cm}{0cm}                    & \rule{0.28cm}{0cm} NLO \rule{0.28cm}{0cm}    & \rule{0.28cm}{0cm} N$^2$LO\rule{0.28cm}{0cm}     &\rule{0.28cm}{0cm} N$^3$LO\rule{0.28cm}{0cm}  &   N$^2$LO \cite{Kang:2013}  &   Experiment       \\
\hline\hline
$E_{^1S_0}$~(eV)       & $-$448  & $-$446 & $-$443  & $-$436  & $-$440(75)~\cite{Augsburger:1999yt}  \\
& & & & & $-$740(150)~\cite{Ziegler:1988bp} \\
$\Gamma_{^1S_0}$~(eV)  &   1155  &  1183  &  1171    &  1174  & 1200(250)~\cite{Augsburger:1999yt}  \\
& & & & &  1600(400)~\cite{Ziegler:1988bp} \\ \hline
$E_{^3S_1}$~(eV)       & $-$742  & $-$766 & $-$770   & $-$756  & $-$785(35)~\cite{Augsburger:1999yt} \\
& & & & & $-$850(42)~\cite{Heitlinger:1991cn} \\
$\Gamma_{^3S_1}$~(eV)  & 1106    &1136    &  1161  &  1120  &  940(80)~\cite{Augsburger:1999yt} \\
& & & & & 770(150)~\cite{Heitlinger:1991cn} \\ \hline
$E_{^3P_0}$~(meV)      & 17      & 12     & 8      & 16     & 139(28)~\cite{Gotta:1999vj}  \\
$\Gamma_{^3P_0}$~(meV) & 194     & 195    &  188   &  169   &  120(25)~\cite{Gotta:1999vj}  \\
\hline\hline
$E_{1S}$~(eV)          & $-$670  & $-$688 & $-$690 & $-$676 & $-$721(14)~\cite{Augsburger:1999yt}  \\
$\Gamma_{1S}$~(eV)     & 1118    &  1148  &  1164  &  1134  & 1097(42)~\cite{Augsburger:1999yt}  \\ \hline
$E_{2P}$~(meV)         & 1.3     & 2.8    & 4.7    & 2.3    & 15(20)~\cite{Gotta:1999vj}  \\
$\Gamma_{2P}$~(meV)    & 36.2    & 37.4   & 37.9   & 27     &  38.0(2.8)~\cite{Gotta:1999vj}  \\
\hline\hline
\end{tabular}
\renewcommand{\arraystretch}{1.0}
\end{center}
\end{table}

\begin{figure}[h!]
\centering
\includegraphics[width=0.45\textwidth,height=0.30\textheight]{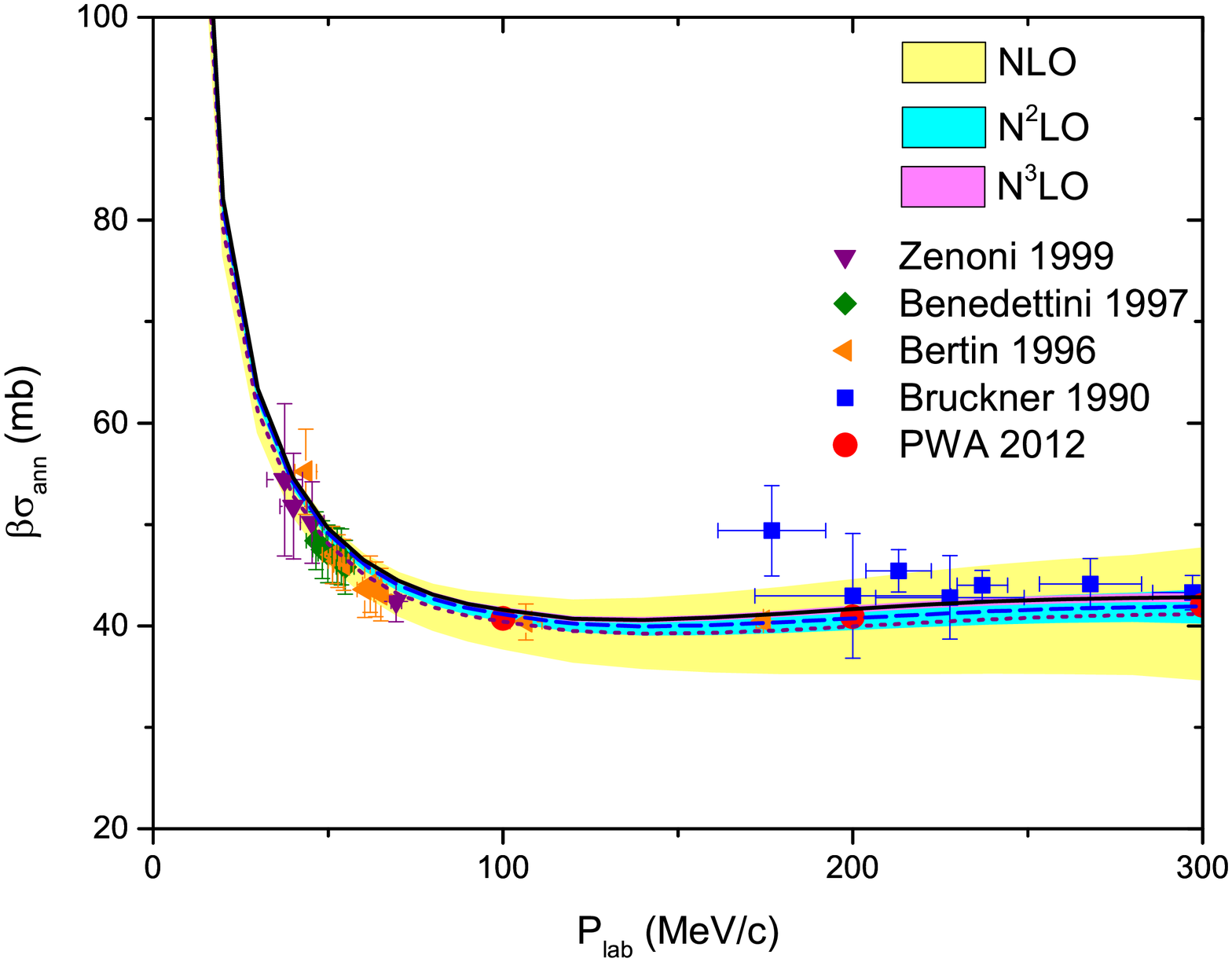}
\includegraphics[width=0.45\textwidth,height=0.30\textheight]{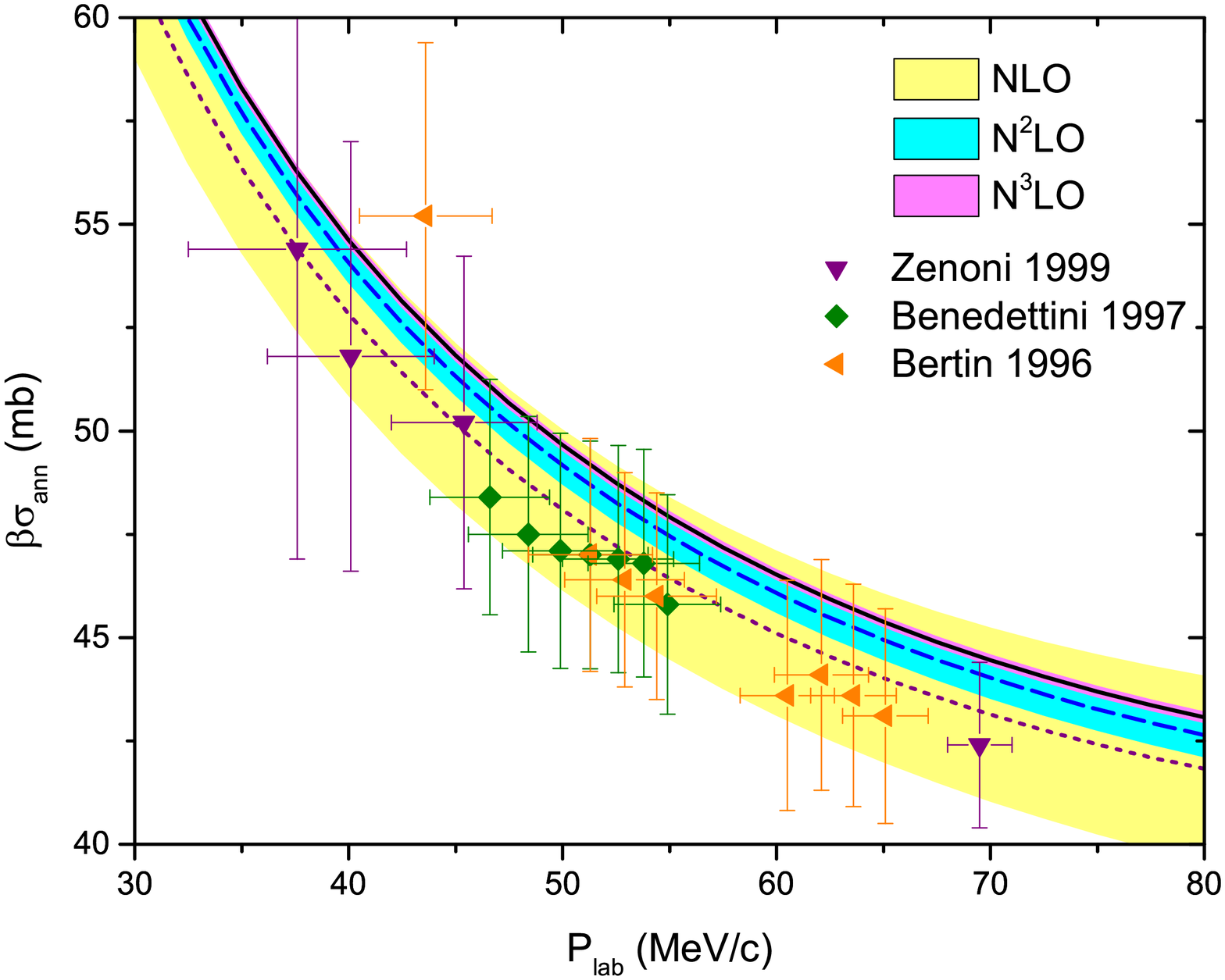}
\caption{$\bar pp$ annihilation cross section multiplied by the velocity $\beta$
of the incoming $\bar p$. For notations, see  Fig.~\ref{fig:cs}.
The results of the PWA \cite{Zhou:2012} are indicated by circles.
Data are taken from \cite{Bruckner1990,Bertin:1996,Benedettini:1997,Zenoni:1999}.
}
\label{fig:ann}
\end{figure}

There are measurements of the $\bar pp$ annihilation cross section
at very low energy \cite{Bruckner1990,Bertin:1996,Benedettini:1997,Zenoni:1999}.
Also those experiments were not taken into account in the PWA \cite{Zhou:2012}.
We present our predictions for this observable in Fig.~\ref{fig:ann}, where
the annihilation cross section multiplied by the velocity $\beta$ of the
incoming $\bar p$ is shown.
Results based on the amplitudes of the PWA are also included (filled circles).
An interesting aspects of those data is that one can see the anomalous
behavior of the reaction cross section near threshold due to the presence of
the attractive Coulomb force \cite{Wigner}. Usually the cross sections for exothermic
reactions behave like $1/\beta$ so that $\beta \sigma_{ann}$ is then practically
constant, cf. Fig.~\ref{fig:ann} for $p_{lab}\approx 100 - 300$~MeV/c.
However, the Coulomb attraction modifies that to a $1/\beta^2$ behavior
for energies very close to the threshold.

\begin{figure}[htbp]
\centering
\includegraphics[width=0.45\textwidth,height=0.30\textheight]{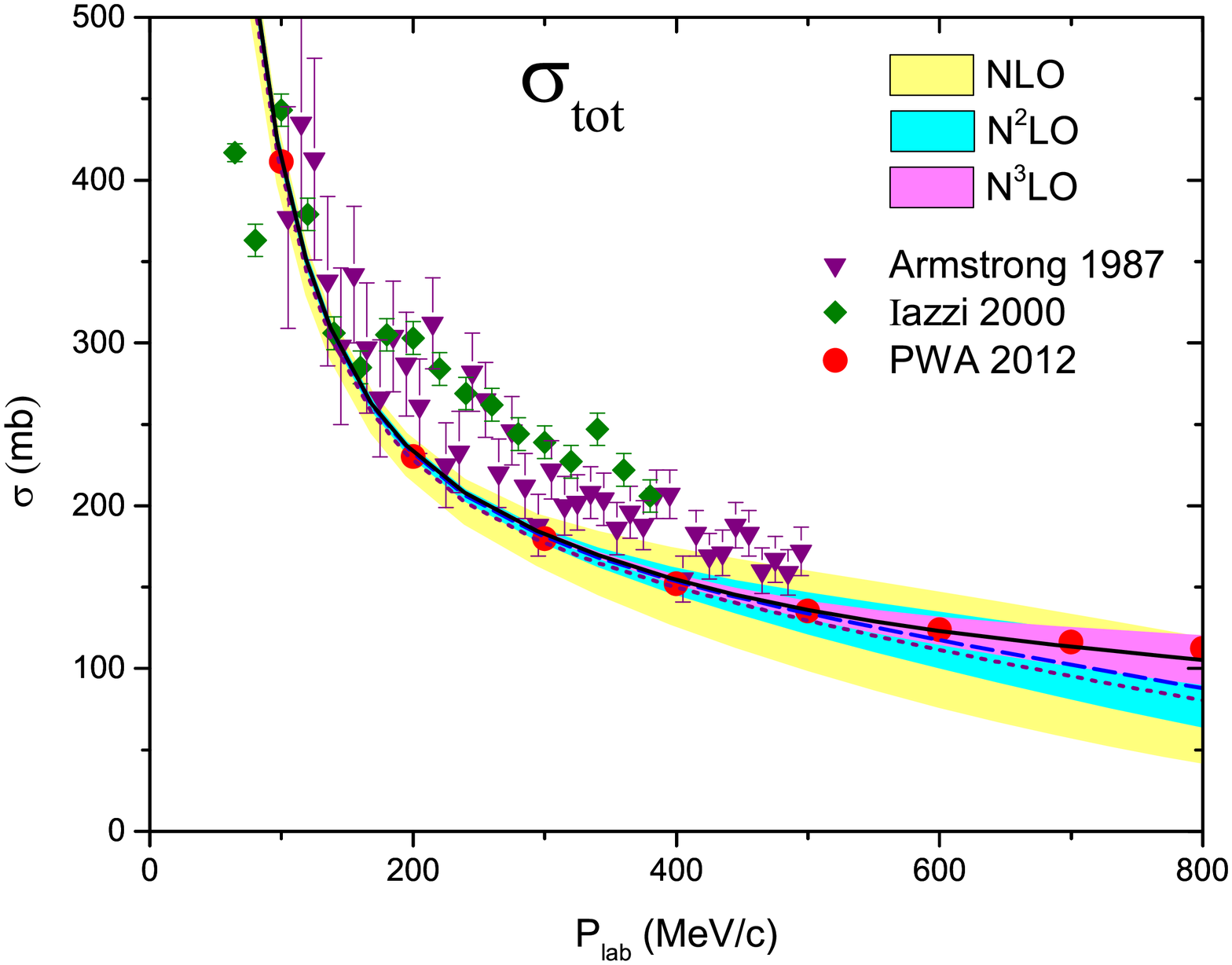}
\includegraphics[width=0.45\textwidth,height=0.30\textheight]{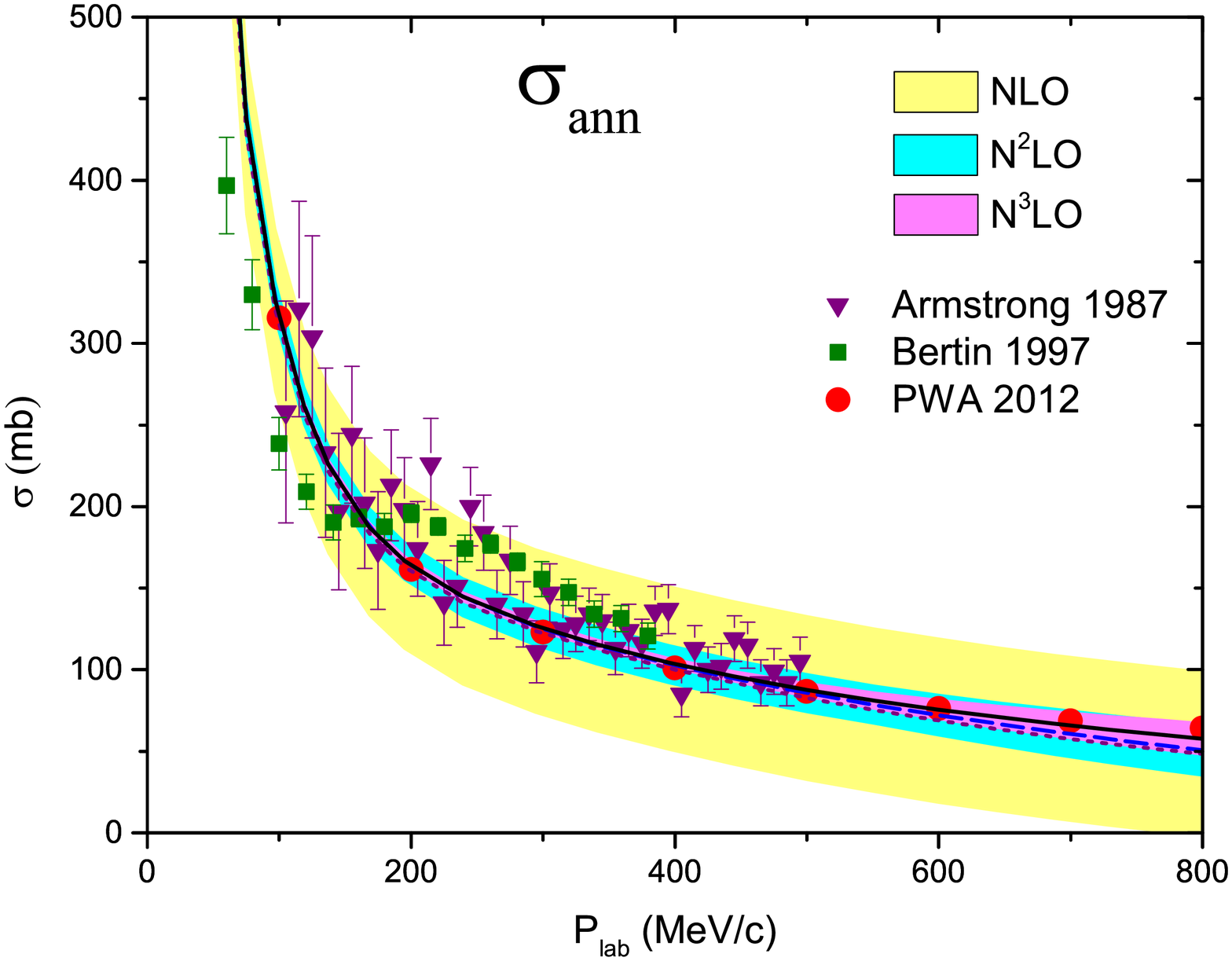}
\caption{Total ($\sigma_{tot}$) and integrated annihilation ($\sigma_{ann}$)
cross sections for $\bar np$ scattering.
For notations, see  Fig.~\ref{fig:cs}.
Data are taken from Refs.~\cite{Arm87,Iaz00,Bertin:1997}.
}
\label{fig:np}
\end{figure}

Finally, for illustration we show our predictions for $\bar np$ scattering, see
Fig.~\ref{fig:np}. The $\bar np$ system is a pure isospin $I=1$ state
so that one can test the $I=1$ component of the $\bar NN$ amplitude
independently.
Note that the PWA results displayed in Fig.~\ref{fig:np} include again partial-wave
amplitudes from our N$^3$LO interaction for $J \geq 5$. However, for integrated
cross sections the contributions of those higher partial waves is really very
small, even at $p_{lab}=800$~MeV/c.

\section{Summary}
\label{sec:5}

In Ref.~\cite{EKM:2015} a new generation of $NN$ potentials derived
in the framework of chiral effective field theory was presented.
In that work a new local regularization scheme was introduced and applied
to the pion-exchange contributions of the $NN$ force. Furthermore, an alternative
scheme for estimating the uncertainty was proposed that no longer depends
on a variation of the cutoffs.
In the present paper we adopted their suggestions and applied them in a study
of the $\bar NN$ interaction. Specifically, a $\bar NN$ potential has been
derived up to N$^3$LO in the perturbative expansion, thereby extending a previous
work by our group that had considered the $\bar NN$ force up to N$^2$LO \cite{Kang:2013}.
Like before, the pertinent low-energy constants have been fixed by a
fit to the phase shifts and inelasticities provided by a recently
published phase-shift analysis of $\bar pp$ scattering data \cite{Zhou:2012}.

We could show that an excellent reproduction of the $\bar NN$ amplitudes
can be achieved at N$^3$LO. Indeed, in many aspects the quality of the description
is comparable to that one has found in case of the $NN$ interaction at the same
order~\cite{EKM:2015}.
To be more specific, for the $S$-waves excellent agreement with the phase shifts and
inelasticities of \cite{Zhou:2012} has been obtained up to laboratory energies of
about $300$~MeV, i.e. over the whole energy range considered. The same is also
the case for most $P$-waves. Even many of the $D$-waves are described well up to
$200$~MeV or beyond.
Because of the overall quality in the reproduction of the individual partial waves
there is also a nice agreement on the level of $\bar NN$ observables. Total
and integrated elastic ($\bar pp\to\bar pp$) and charge-exchange ($\bar pp\to\bar nn$)
cross sections agree well with the PWA results up to the highest energy considered
while differential observables (cross sections, analyzing powers, etc.) are
reproduced quantitatively up to $200$-$250$ MeV.
Furthermore, and equally important, in most of the considered cases the achieved
results agree with the ones based on the PWA within the estimated theoretical accuracy.
Thus, the scheme for quantifying the uncertainty suggested in Ref.~\cite{EKM:2015}
seems to work well and can be applied reliably to the $\bar NN$ interaction
as well.
Finally, the low-energy representation of the $\bar NN$ amplitudes
derived from chiral EFT compares well with the constraints derived from the phenomenology
of antiprotonic hydrogen.

\section*{Acknowledgements}

We would like to thank Evgeny Epelbaum for useful discussions.
We are also grateful to Detlev Gotta for clarifying discussions on
various issues related to the measurements of antiprotonic hydrogen.
This work is supported in part by the DFG and the NSFC through
funds provided to the Sino-German CRC 110 ``Symmetries and
the Emergence of Structure in QCD'' and the BMBF (contract No. 05P2015 -NUSTAR R\&D).
The work of UGM was supported in part by The Chinese Academy
of Sciences (CAS) President's International Fellowship Initiative (PIFI) grant no.~2017VMA0025.

\appendix
\setcounter{equation}{0}
\setcounter{table}{0}
\renewcommand{\theequation}{\Alph{section}.\arabic{equation}}
\renewcommand{\thetable}{\Alph{section}.\arabic{table}}

\section{The chiral potential up to N$^3$LO}\label{app:potential}
The one-pion exchange potential (OPEP) is given in Eq.~(\ref{opep_full}).
Up to N$^3$LO, the chiral expansion of the two-pion exchange potential (TPEP) can be found
in Refs.~\cite{Epe05,EKM:2015,Kaiser:2001pc}. For the reader's convenience
we summarize the expressions below. The TPEP can be written in the form
\beqa
\label{2PEdec}
V_{2 \pi}  &=& V_C + \fet \tau_1 \cdot \fet \tau_2 \, W_C + \left[
V_S + \fet \tau_1 \cdot \fet \tau_2 \, W_S \right] \, \fet \sigma_1 \cdot \fet \sigma_2
+ \left[ V_T + \fet \tau_1 \cdot \fet \tau_2 \, W_T \right]
\, \fet \sigma_1 \cdot \fet q \, \fet \sigma_2 \cdot \fet q \nn
&+& \left[ V_{LS} + \fet \tau_1 \cdot \fet \tau_2 \, W_{LS} \right]
\, i (\fet \sigma_1 + \fet \sigma_2 ) \cdot ( \fet q \times \fet k ) \,,
\eeqa
where ${\fet q} = {\fet p}' - {\fet p}$, ${\fet k} = ({\fet p}' + {\fet p})/2$,
and $\fet \tau_i$ is the isospin Pauli matrix associated
with the nucleon (antinucleon) $i$.
$V$ denotes the isoscalar part and $W$ the isovector part where the subscripts $C$, $S$, $T$, $LS$
refer to the central, spin-spin, tensor, and spin-orbit terms, respectively.
Each component of $V$ and $W$ is given by a sum $V=V^{(0)}+V^{(2)}+V^{(3)}+V^{(4)}$ (analogous for $W$)
where the superscript in the bracket refers to the chiral dimension.
The order-$Q^2$ contributions take the form
\beqa
\label{2PE_nlo}
W_C^{(2)} &=& - \frac{L (q )}{384 \pi^2 F_\pi^4}
\bigg[4M_\pi^2 (5g_A^4 - 4g_A^2 -1)  + q^2(23g_A^4 - 10g_A^2 -1)
+ \frac{48 g_A^4 M_\pi^4}{4 M_\pi^2 + q^2} \bigg] \,, \nn
V_T^{(2)} &=& -\frac{1}{q^2} V_S^{(2)}  = - \frac{3 g_A^4}{64
  \pi^2 F_\pi^4} \,L (q)\,,\nn
V_C^{(2)} &=& V_{LS}^{(2)} = W_S^{(2)} = W_T^{(2)} = W_{LS}^{(2)} = 0\,.
\eeqa
The loop function $L ( q)$ is defined in dimensional regularization (DR) via
\beq
\label{def_LA}
L ( q) = \frac{\sqrt{4 M_\pi^2 + q^2}}{q} \ln
\frac{\sqrt{4 M_\pi^2 + q^2} + q}{2 M_\pi} \,.
\eeq
Notice that all polynomial terms are absorbed into contact interactions, as given in Eqs.~(\ref{C1S0})-(\ref{VC}).
The corrections at order $Q^3$ giving rise to the subleading TPEP have the form
\beqa
\label{2PE_nnlo}
V_C^{(3)}  &=& -\frac{3g_A^2}{16\pi F_\pi^4}  \bigg[2M_\pi^2(2c_1 -c_3) -c_3 q^2 \bigg]
(2M_\pi^2+q^2) A ( q  )\,, \nn
W_T^{(3)} &=& -\frac{1}{q^2} W_S^{(3)}  = - \frac{g_A^2}{32\pi F_\pi^4} \,  c_4 (4M_\pi^2 + q^2)
A( q )\,, \nn
V_S^{(3)} &=& V_{T}^{(3)} = V_{LS}^{(3)} = W_C^{(3)} = W_{LS}^{(3)} = 0\,,
\eeqa
where the loop function $A ( q )$ is given in DR by
\beq
A ( q  ) = \frac{1}{2  q } \arctan \frac{ q }{2 M_\pi} \,.
\eeq
At order $Q^4$, i.e. N$^3$LO, the contributions of one-loop ``bubble'' diagrams to the TPEP
are
\beqa
\label{2PE_nnnlo}
V_C^{(4)} &=& \frac{3}{16 \pi^2  F_\pi^4} \, L (q) \,
\left\{ \left[ \frac{c_2}{6} (4M_\pi^2 + q^2) + c_3 (2 M_\pi^2 + q^2 ) - 4 c_1 M_\pi^2
\right]^2 + \frac{c_2^2}{45} (4M_\pi^2 + q^2)^2 \right\}\,,
\nonumber \\
W_T^{(4)} &=& -\frac{1}{q^2} W_S^{(4)}= \frac{c_4^2}{96 \pi^2 F_\pi^4} (4M_\pi^2 + q^2)
\, L(q)\,.
\eeqa

Since the regularization is done in coordinate space the potentials have to
be Fourier transformed. For the contributions above this can be done analytically and
the corresponding expressions (up to N$^2$LO) have been given in~\cite{Kaiser1997,Gezerlis:2014}.
\begin{eqnarray}
\noindent W_S^{(0)}(r)&=& \frac{g_A^2 x^2 e^{-x}}{48 \pi  F_\pi^2 r^3}\;,\\
\noindent \tilde W_T^{(0)}(r)&=&\frac{g_A^2 e^{-x} (x^2+3 x+3)}{48 \pi  F_\pi^2 r^3}\;,\\
\noindent V_S^{(2)}(r)&=&\frac{g_A^4 x \left((2 x^2+3) K_1(2 x)+3 x K_0(2 x)\right)}{32
\pi ^3 F_\pi^4 r^5}\;,\\
\noindent \tilde V_T^{(2)}(r)&=&-\frac{g_A^4 x \left((4 x^2+15) K_1(2 x)+12 x K_0(2 x)\right)}{128
\pi ^3 F_\pi^4 r^5}\;,\\
\noindent W_C^{(2)}(r)&=&-\frac{x \left(x \left(g_A^4 (4 x^2+23)-10 g_A^2-1\right) K_0(2 x)+\left(g_A^4
(12 x^2+23)-2 g_A^2 (2 x^2+5)-1\right) K_1(2 x)\right)}{128 \pi ^3 F_\pi^4
r^5}\;,\\
\noindent V_C^{(3)}(r)&=&\frac{3 g_A^2 e^{-2 x} \left(2 c_1 x^2 (x+1)^2+c_3 (x^4+4 x^3+10 x^2+12 x+6)\right)}{32 \pi ^2 F_\pi^4 r^6}\;,\\
\noindent W_S^{(3)}(r)&=&\frac{c_4 g_A^2 e^{-2 x} (x+1) (2 x^2+3 x+3)}{48 \pi ^2 F_\pi^4
r^6}\;,\\
\noindent \tilde W_T^{(3)}(r)&=&-\frac{c_4 g_A^2 e^{-2 x} (x+1) (x^2+3 x+3)}{48 \pi ^2 F_\pi^4
r^6}\;,
\end{eqnarray}
where $x=M_{\pi}r$, $K_i(x)$ is the modified Bessel function of the second kind and the superscript
in the bracket refers to the chiral dimension. Note that the tensor parts of the potentials in coordinate
space ($\tilde V_T$, $\tilde W_T$) are written with a tilde as a reminder that they are defined in terms
of the irreducible tensor operator
$S_{12} = 3(\fet \sigma_1 \cdot \fet {\hat r} \, \fet \sigma_2 \cdot \fet {\hat r}) - \fet \sigma_1 \cdot \fet \sigma_2$
where $\fet {\hat r} = \fet r / r$.

The relativistic, i.e. the $1/m$,  corrections are given by
\begin{eqnarray}
\noindent V_{C,m}^{(4)}(r)&=&\frac{3 g_A^4 e^{-2 x} (x^5+10 x^4+28 x^3+46 x^2+48 x+24)}{1024 \pi ^2 F_\pi^4 m r^6}\;,\\
\noindent V_{S,m}^{(4)}(r)&=&-\frac{g_A^4 e^{-2 x} (6 x^4+22 x^3+43 x^2+48 x+24)}{512
\pi ^2 F_\pi^4 m r^6}\;,\\
\noindent \tilde V_{T,m}^{(4)}(r)&=&\frac{g_A^4 e^{-2 x} (6 x^4+31 x^3+76 x^2+96 x+48)}{1024
\pi ^2 F_\pi^4 m r^6}\;,\\
\noindent V_{LS,m}^{(4)}(r)&=&-\frac{3 g_A^4 e^{-2 x} (x+1) (x^2+2 x+2)}{64 \pi ^2 F_\pi^4 m
r^6}\;,\\
\noindent W_{C,m}^{(4)}(r)&=&\frac{g_A^2 e^{-2 x} \left(g_A^2 (3 x^5+10 x^4+36 x^3+82 x^2+96 x+48)-4(x^4+4 x^3+10 x^2+12 x+6)\right)}{512 \pi ^2 F_\pi^4 m r^6}\;,\\
\noindent W_{S,m}^{(4)}(r)&=&-\frac{g_A^2 e^{-2 x} \left(g_A^2 (2 x^4+10 x^3+21 x^2+24 x+12)-4(2 x^3+5 x^2+6 x+3)\right)}{768 \pi ^2 F_\pi^4 m r^6}\;,\\
\noindent \tilde W_{T,m}^{(4)}(r)&=&\frac{g_A^2 e^{-2 x} \left(g_A^2 (2 x^4+13 x^3+36 x^2+48 x+24)-8(x^3+4 x^2+6 x+3)\right)}{1536 \pi ^2 F_\pi^4 m r^6}\;,\\
\noindent W_{LS,m}^{(4)}(r)&=&\frac{g_A^2 (g_A^2-1) e^{-2 x} (x+1)^2}{32 \pi ^2 F_\pi^4 m r^6}\;.
\end{eqnarray}
The subleading order corrections to the $\pi N$ vertex are given by
\begin{eqnarray}
\noindent V_{C,sl}^{(4)}(r)&=&-\frac{g_A^2 x}{128 \pi ^3 F_\pi^4 m r^7} \left[K_0(2 x) \left(48 x^3 (6 c_1+c_2-3
c_3)+24 x^5 (2 c_1+c_3)-48 x (c_2-6 c_3)(2 x^2+5)\right)\right. \nonumber\\
&&-2 K_1(2 x) \left(-16 x^4 (6 c_1+c_2-3 c_3)-24 x^2 (6 c_1+c_2-3 c_3)+6 x^4 (c_2-2
c_3)\right.\nonumber\\
&&+\left.\left.4 (c_2-6 c_3) (4 x^4+27 x^2+30)\right)\right]\;,\\
\noindent V_{LS,sl}^{(4)}(r)&=&\frac{3 c_2 g_A^2 x \left( (2 x^2+5) K_1(2 x)+5 x K_0(2 x)\right)}{8\pi ^3 F_\pi^4 m r^7}\;,\\
\noindent W_{C,sl}^{(4)}(r)&=&\frac{c_4 x \left(2 x \left(g_A^2 (8 x^2+25)+x^2+5\right) K_0(2 x)+\left(g_A^2 (4 x^4+41 x^2+50)+7 x^2+10\right) K_1(2 x)\right)}{32 \pi ^3 F_\pi^4 m r^7}\;,\\
\noindent W_{S,sl}^{(4)}(r)&=&-\frac{c_4 x \left(x \left(g_A^2 (4 x^2+35)-5\right) K_1(2 x)+2 \left(5
g_A^2 (2 x^2+7)-x^2-5\right) \left(x K_0(2 x)+K_1(2 x)\right)\right)}{48 \pi ^3 F_\pi^4 m r^7}\;,\\
\noindent \tilde W_{T,sl}^{(4)}(r)&=&\frac{c_4 x \left(2 x \left(g_A^2(4 x^2+59)-8\right) K_1(2 x)+\left(g_A^2
(52 x^2+245)-4 x^2-35\right) \left(x K_0(2 x)+K_1(2 x)\right)\right)}{192\pi ^3 F_\pi^4 m r^7}\;,\\
\noindent W_{LS,sl}^{(4)}(r)&=&-\frac{c_4 x \left(x \left(g_A^2 (4 x^2+25)+5\right) K_0(2 x)+\left(g_A^2
(16 x^2+25)+2 x^2+5\right) K_1(2 x)\right)}{16 \pi ^3 F_\pi^4 m r^7}\;.
\end{eqnarray}
The one loop \lq bubble' diagrams corrections to the TPEP potential amount to
\begin{eqnarray}
\noindent V_{C,b}^{(4)}(r)&=&-\frac{3x}{32 \pi ^3 F_\pi^4 r^7}\left[K_1(2 x) \left(4 \left(4 c_1^2 x^4+4 c_1 c_3
x^2 (x^2+3)+c_3^2 (x^4+21 x^2+30)\right) \right.\right.\nonumber\\
&&\left.+8 c_2 \left(c_1x^2+c_3 (3 x^2+5)\right)+3 c_2^2 (x^2+2)\right)\nonumber\\
&&\left.+2 x K_0(2 x) \left(2 c_2 (2 c_1 x^2+c_3 x^2+10 c_3)+12 c_3 (2 c_1 x^2+c_3 x^2+5 c_3)+3 c_2^2\right)\right]\;,\\
\noindent W_{S,b}^{(4)}(r)&=&\frac{c_4^2 x \left(2 x (x^2+5) K_0(2 x)+(7 x^2+10)K_1(2 x)\right)}{24 \pi ^3 F_\pi^4 r^7}\;,\\
\noindent \tilde W_{T,b}^{(4)}(r)&=&-\frac{c_4^2 x \left(x (4 x^2+35) K_0(2 x)+5(4 x^2+7)K_1(2 x)\right)}{96 \pi ^3 F_\pi^4 r^7}\;.
\end{eqnarray}

There are further contributions to the TPEP at N$^3$LO where one cannot get analytical forms in
coordinate space.
Most conveniently one can write those in the (subtracted) spectral representation
\beqa
V_{C,S} (q) &=& - \frac{2 q^6}{\pi} \int_{2M_\pi}^\infty \, d \mu
\frac{\rho_{C,S} (\mu )}{\mu^5 ( \mu^2 + q^2 )}\,, \quad \quad
V_T (q) = \frac{2 q^4}{\pi} \int_{2M_\pi}^\infty \, d \mu
\frac{\rho_{T} (\mu )}{\mu^3 ( \mu^2 + q^2 )}\,, \nn
W_{C,S} (q) &=& - \frac{2 q^6}{\pi} \int_{2M_\pi}^\infty \, d \mu
\frac{\eta_{C,S} (\mu )}{\mu^5 ( \mu^2 + q^2 )}\,, \quad \quad
W_T (q) = \frac{2 q^4}{\pi} \int_{2M_\pi}^\infty \, d \mu
\frac{\eta_{T} (\mu )}{\mu^3 ( \mu^2 + q^2 )}\,,
\eeqa
where $\rho_i$ and $\eta_i$ denote the corresponding spectral functions which are related to the
potential via $\rho_i (\mu) = {\rm Im} V_i (i \mu )$,  $\eta_i (\mu) = {\rm Im} W_i (i \mu )$.
For the spectral functions $\rho_i (\mu)$ ($\eta_i (\mu)$) one finds
\cite{Kaiser:2001pc}:
\beqa
\label{TPE2loop}
\rho_C^{(4)} (\mu ) &=& - \frac{3 g_A^4 (\mu^2 - 2 M_\pi^2 )}{\pi \mu (4 F_\pi)^6}
\,
\bigg\{ (M_\pi^2 - 2 \mu^2 ) \bigg[ 2 M_\pi + \frac{2 M_\pi^2 - \mu^2}{2 \mu}
\ln \frac{\mu + 2 M_\pi}{\mu - 2 M_\pi } \bigg] + 4 g_A^2 M_\pi (2 M_\pi^2 - \mu^2 )
\bigg\}\,, \nn
\eta_S^{(4)} (\mu ) &=& \mu^2 \eta_T^{(4)} (\mu ) = -
\frac{g_A^4 (\mu^2 - 4 M_\pi^2 )}{\pi (4 F_\pi)^6}
\,
\left\{ \left(M_\pi^2 - \frac{\mu^2}{4} \right) \ln \frac{\mu +  2 M_\pi}{\mu - 2 M_\pi }
+ (1 + 2 g_A^2 ) \mu M_\pi \right\}\,, \nn
\rho_S^{(4)} (\mu ) &=& \mu^2 \rho_T^{(4)} (\mu ) = -
  \left\{
\frac{g_A^2 r^3 \mu}{8 F_\pi^4 \pi}
(\bar d_{14} - \bar d_{15} ) -
\frac{2 g_A^6 \mu r^3}{(8 \pi F_\pi^2)^3} \left[ \frac{1}{9} - J_1  + J_2 \right] \right\}\,, \nn
 \eta_C^{(4)} (\mu ) &=& \Bigg\{
\frac{r t^2}{24 F_\pi^4 \mu \pi} \left[ 2 (g_A^2 - 1) r^2 - 3 g_A^2 t^2 \right] (\bar d_1 + \bar d_2 ) \nn
&& {}+ \frac{r^3}{60 F_\pi^4 \mu \pi} \left[ 6 (g_A^2 - 1) r^2 - 5 g_A^2 t^2 \right] \bar d_3
- \frac{r M_\pi^2}{6 F_\pi^4 \mu \pi} \left[ 2 (g_A^2 - 1) r^2 - 3 g_A^2 t^2 \right] \bar d_5 \nn
&& {} - \frac{1}{92160 F_\pi^6 \mu^2 \pi^3} \Big[ - 320 (1 + 2 g_A^2 )^2 M_\pi^6 +
240 (1 + 6 g_A^2 + 8 g_A^4 ) M_\pi^4 \mu^2 \nn
&& {}   \mbox{\hskip 3 true cm} - 60 g_A^2 (8 + 15 g_A^2 ) M_\pi^2  \mu^4
+ (-4 + 29 g_A^2 + 122 g_A^4 + 3 g_A^6 ) \mu^6 \Big] \ln \frac{2 r + \mu}{2 M_\pi} \nn
&& {} - \frac{r}{2700 \mu ( 8 \pi F_\pi^2 )^3} \Big[ -16 ( 171 + 2 g_A^2 ( 1 + g_A^2)
(327 + 49 g_A^2)) M_\pi^4 \nn
&& {}   \mbox{\hskip 3 true cm} + 4 (-73 + 1748 g_A^2 + 2549 g_A^4 + 726 g_A^6 ) M_\pi^2 \mu^2 \nn
&& {}   \mbox{\hskip 3 true cm} - (- 64 + 389 g_A^2 + 1782 g_A^4 + 1093 g_A^6 ) \mu^4  \Big] \nn
&& {} + \frac{2 r}{3 \mu ( 8 \pi F_\pi^2 )^3} \Big[
g_A^6 t^4 J_1 - 2 g_A^4 (2 g_A^2 -1 ) r^2 t^2 J_2 \Big] \Bigg\}\,,
\eeqa
where the abbreviations are
\beq
r = \frac{1}{2} \sqrt{ \mu^2  - 4 M_\pi^2}\,, \quad \quad \quad
t= \sqrt{\mu^2 - 2 M_\pi^2}\,,
\eeq
and
\beqa
J_1 &=& \int_0^1 \, dx \, \bigg\{ \frac{M_\pi^2}{r^2 x^2} - \bigg( 1 + \frac{M_\pi^2 }{r^2 x^2} \bigg)^{3/2}
\ln \frac{ r x + \sqrt{ M_\pi^2 + r^2 x^2}}{M_\pi} \bigg\}\,,\nonumber \\
J_2 &=& \int_0^1 \, dx \, x^2 \bigg\{ \frac{M_\pi^2}{r^2 x^2} - \bigg( 1 + \frac{M_\pi^2 }{r^2 x^2} \bigg)^{3/2}
\ln \frac{ r x + \sqrt{ M_\pi^2 + r^2 x^2}}{M_\pi} \bigg\}\,.
\eeqa
The LECs $\bar d_{1}, \; \bar d_{2}, \; \bar d_{3}, \; \bar d_{5}, \; \bar d_{14}$ and $\bar d_{15}$ are discussed in Sec.~\ref{sec:2MEX}.
This part of the potential is Fourier transformed numerically.

After regularization in coordinate space, we need to transform the potential back to momentum
space where we solve the Lippmann-Schwinger equation. For that we employ the master formulae given
in Ref.~\cite{Veerasamy2011} and obtain:
\begin{eqnarray}
V_C(q)&=&4 \pi  \int _0^{\infty } f(r)\, V_C(r) j_0(q r)r^2dr\;,\nonumber\\
V_S(q)&=&4 \pi  \int _0^{\infty } f(r)\, \left(V_S(r) j_0(q r)+\tilde V_T(r) j_2(q r)\right) r^2dr\;,\nonumber\\
V_T(q)&=&-\frac{12 \pi  }{q^2}\int _0^{\infty } f(r)\, \tilde V_T(r) j_2(q r)r^2dr\;,\nonumber\\
V_{SL}(q)&=&\frac{4 \pi  }{q}\int _0^{\infty }f(r)\, V_{LS}(r) j_1(q r)r^3 dr\; ,
\end{eqnarray}
Here $f(r)$ is the regulator function given in Eq.~(\ref{cutoff}).
The same relations apply also to the isovector part $W$.

\section{Values of the low-energy constants}
\label{app:LECs}
Values of the LECs obtained in our fit to the $\bar NN$ phase shifts of the
PWA \cite{Zhou:2012} at N$^3$LO are collected in Tables~\ref{tab:cutoff}
and \ref{tab:cutoff1} for the cutoffs $R=0.8-1.2$~fm.
Values of the LECs obtained in our fit to the $\bar NN$ phase shifts
of the PWA \cite{Zhou:2012} at LO, NLO, N$^2$LO for the cutoffs $R=0.9$~fm and $R=1.0$~fm
are collected in Table~\ref{tab:order}.

\begin{table}[htbp]
\begin{center}
\vspace{-1.5cm}
\begin{tabular}{|c||c|c|c|c|c|c|}
\hline\hline
\rule[0.5cm]{0.2cm}{0cm}LEC\rule[0.5cm]{0.2cm}{0cm}  & \rule{0.28cm}{0cm} R=0.8\,fm \rule{0.28cm}{0cm}    & \rule{0.28cm}{0cm}R=0.9\,fm\rule{0.28cm}{0cm}
&\rule{0.28cm}{0cm} R=1.0\,fm \rule{0.28cm}{0cm}       &\rule{0.28cm}{0cm} R=1.1\,fm \rule{0.28cm}{0cm}     & \rule{0.28cm}{0cm}R=1.2\,fm\rule{0.28cm}{0cm}      \\
\hline\hline
$\tilde{C}_{^{11}S_0}$~(GeV$^{-2}$)  &   0.1816     &  0.0734      &   0.0293       &   0.0007       &    -0.0089    \\
$C_{^{11}S_0}$~(GeV$^{-4}$)          &  -0.2134     &  0.0032      &   0.1353       &   0.1754       &     0.2188     \\
$D^1_{^{11}S_0}$~(GeV$^{-6}$)        &  -2.8614     & -3.8012      &  -4.5883       &  -3.7943       &    -0.0746    \\
$D^2_{^{11}S_0}$~(GeV$^{-8}$)        &   2.1256     &  2.2443      &   3.0715       &   4.5639       &     6.4500  \\
$\tilde{C}^a_{^{11}S_0}$~(GeV$^{-1}$)&  -0.5809     & -0.5437      &  -0.5326       &  -0.5173       &    -0.5007    \\
$C^a_{^{11}S_0}$~(GeV$^{-3}$)        &   0.5993     &  0.1067      &  -0.5747       &  -1.5894       &    -2.7102     \\
\hline
$\tilde{C}_{^{31}S_0}$~(GeV$^{-2}$)  &    0.3960    &  0.1870      &   0.1189       &   0.0889       &     0.0737    \\
$C_{^{31}S_0}$~(GeV$^{-4}$)          &    0.0966    & -0.0702      &  -0.0796       &  -0.1356       &    -0.1228     \\
$D^1_{^{31}S_0}$~(GeV$^{-6}$)        &    2.8886    &  1.4143      &  -1.3774       &  -5.0372       &    -9.9353    \\
$D^2_{^{31}S_0}$~(GeV$^{-8}$)        &    1.4470    &  1.7541      &   3.2624       &   6.6055       &     9.9568  \\
$\tilde{C}^a_{^{31}S_0}$~(GeV$^{-1}$)&   -0.5768    & -0.5102      &  -0.5078       &  -0.5251       &    -0.5293    \\
$C^a_{^{31}S_0}$~(GeV$^{-3}$)        &    0.4809    &  0.3750      &   0.1227       &  -0.3239       &    -0.7031     \\
\hline
$\tilde{C}_{^{13}S_1}$~(GeV$^{-2}$)  &    1.6589    &  0.8795      &   0.5625       &   0.3402       &     0.2346    \\
$C_{^{13}S_1}$~(GeV$^{-4}$)          &   -1.4947    & -1.3232      &  -1.2473       &  -1.2201       &    -1.3363     \\
$D^1_{^{13}S_1}$~(GeV$^{-6}$)        &   -1.0563    & -4.1331      &  -8.2720       & -13.3684       &   -22.8316    \\
$D^1_{^{13}S_1}$~(GeV$^{-8}$)        &   -2.3730    & -5.0615      &  -8.3651       & -13.4933       &   -17.9644  \\
$\tilde{C}^a_{^{13}S_1}$~(GeV$^{-1}$)&   -1.1612    & -0.9880      &  -0.9724       &  -1.0715       &    -1.0846    \\
$C^a_{^{13}S_1}$~(GeV$^{-3}$)        &    1.4455    &  1.8999      &   2.8473       &   4.0483       &     5.3069     \\
\hline
$\tilde{C}_{^{33}S_1}$~(GeV$^{-2}$)  &    0.2214    &  0.2537      &   0.2621       &   0.1740       &     0.0984    \\
$C_{^{33}S_1}$~(GeV$^{-4}$)          &   -0.8849    & -0.7500      &  -0.1184       &  -0.0442       &    -0.0864     \\
$D^1_{^{33}S_1}$~(GeV$^{-6}$)        &   -0.9113    & -2.5135      &  -3.1696       &  -2.5085       &    -0.2544     \\
$D^2_{^{33}S_1}$~(GeV$^{-8}$)        &    5.8826    &  7.0499      &   8.2400       &   9.5785       &    10.9252 \\
$\tilde{C}^a_{^{33}S_1}$~(GeV$^{-1}$)&    1.7798    &  1.0938      &   0.7817       &   0.6102       &     0.5040    \\
$C^a_{^{33}S_1}$~(GeV$^{-3}$)        &    1.6053    &  2.0396      &   2.0031       &   2.3243       &     2.9964     \\
\hline
$C_{^{1}\epsilon_1}$~(GeV$^{-4}$)    &   -1.2873    & -1.0422      &  -1.0352       &  -1.1118       &    -1.2042    \\
$D^1_{^{1}\epsilon_1}$~(GeV$^{-6}$)  &    1.5672    &  2.4207      &   3.1532       &   4.1075       &     4.9037     \\
$D^2_{^{1}\epsilon_1}$~(GeV$^{-8}$)  &    8.9117    &  9.0537      &  10.8574       &  14.7047       &    17.9407    \\
$C^a_{^{1}\epsilon_1}$~(GeV$^{-3}$)  &-0.1132       & -0.3203      &  -0.7550       &  -1.1708       &    -1.7234    \\
\hline
$C_{^{3}\epsilon_1}$~(GeV$^{-4}$)    &-0.8700      & -0.3729      &  -0.0271       &   0.1361       &     0.4158    \\
$D^1_{^{3}\epsilon_1}$~(GeV$^{-6}$)  &0.0661       & -0.4703      &  -2.3147       &  -6.4892       &    -9.1563     \\
$D^2_{^{3}\epsilon_1}$~(GeV$^{-8}$)  & 9.9717      &  9.9728      &   9.1269       &   9.9530       &     9.9987    \\
$C^a_{^{3}\epsilon_1}$~(GeV$^{-3}$)  & 0.5098      &  0.2399      &  -0.3413       &  -0.9511       &    -1.1465    \\
\hline
$C_{^{13}P_0}$~(GeV$^{-4}$)         &   -0.7131    & -1.2874      &  -1.7249       &  -2.1000       &     -2.9288    \\
$D^1_{^{13}P_0}$~(GeV$^{-6}$)       &    0.8404    &  0.4728      &  -1.3347       &  -5.4425       &     -9.5526     \\
$C^a_{^{13}P_0}$~(GeV$^{-2}$)       &   -0.5149    & -0.4760      &  -0.4338       &  -0.3411       &     -0.6103    \\
$D^a_{^{13}P_0}$~(GeV$^{-4}$)       &   -1.4175    & -2.5931      &  -4.2633       &  -7.0558       &     -8.8683     \\
\hline
$C_{^{33}P_0}$~(GeV$^{-4}$)         &   -0.2927    & -0.2364      &  -0.2263       &  -0.0486       &      0.1721    \\
$D^1_{^{33}P_0}$~(GeV$^{-6}$)       &   -0.4391    & -1.8988      &  -3.6350       &  -6.7300       &     -10.8920    \\
$C^a_{^{33}P_0}$~(GeV$^{-2}$)       &    0.4600    &  0.4023      &   0.4034       &   0.3466       &      0.2792    \\
$D^a_{^{33}P_0}$~(GeV$^{-4}$)       &    0.6467    &  1.8980      &   3.5552       &   6.1373       &      9.6870     \\
\hline
$C_{^{11}P_1}$~(GeV$^{-4}$)        &    0.3393    &  0.3290      &   0.3235       &   0.1592       &     -0.0480    \\
$D^1_{^{11}P_1}$~(GeV$^{-6}$)      &    0.2221    &  1.2951      &   2.7015       &   5.1788       &      8.2461     \\
$C^a_{^{11}P_1}$~(GeV$^{-2}$)      &   -1.0467    & -1.0598      &  -1.0268       &  -1.0098       &     -0.9880    \\
$D^a_{^{11}P_1}$~(GeV$^{-4}$)      &   -0.7349    & -1.6834      &  -2.5470       &  -3.5875       &     -4.7692     \\
\hline\hline
\end{tabular}
\caption{\label{tab:cutoff} Low-energy constants at N$^3$LO for different cutoffs.
The superscript $a$ indicates parameters that are related to the annihilation part,
cf. Eqs.~(17)-(\ref{ANN3}).
Note that all parameters are in units of $10^4$.
}
\end{center}
\end{table}

\begin{table}[htbp]
\begin{center}
\vspace{-1.5cm}
\begin{tabular}{|c||c|c|c|c|c|c|}
\hline\hline
\rule[0.5cm]{0.2cm}{0cm}LEC\rule[0.5cm]{0.2cm}{0cm}  & \rule{0.28cm}{0cm} R=0.8\,fm \rule{0.28cm}{0cm}    & \rule{0.28cm}{0cm}R=0.9\,fm\rule{0.28cm}{0cm}
&\rule{0.28cm}{0cm} R=1.0\,fm \rule{0.28cm}{0cm}       &\rule{0.28cm}{0cm} R=1.1\,fm \rule{0.28cm}{0cm}     & \rule{0.28cm}{0cm}R=1.2\,fm\rule{0.28cm}{0cm}      \\
\hline\hline
$C_{^{31}P_1}$~(GeV$^{-4}$)       &    0.4044    &  0.2913      &   0.2541       &   0.0031       &     -0.2178    \\
$D^1_{^{31}P_1}$~(GeV$^{-6}$)     &   0.7746     &  2.1280      &   3.4240       &   5.9236       &      8.7283     \\
$C^a_{^{31}P_1}$~(GeV$^{-2}$)     &   -1.1010    & -1.1208      &  -1.0835       &  -1.0868       &     -1.0770    \\
$D^a_{^{31}P_1}$~(GeV$^{-4}$)     &   -0.8621    & -1.6859      &  -2.2918       &  -3.2685       &     -4.4792     \\
\hline
$C_{^{13}P_1}$~(GeV$^{-4}$)       &   -0.0481    &  0.0110      &   0.1200       &   0.3533       &      0.5733    \\
$D^1_{^{13}P_1}$~(GeV$^{-6}$)     &   -0.1271    & -0.7636      &  -1.5221       &  -2.8235       &     -3.8502     \\
$C^a_{^{13}P_1}$~(GeV$^{-2}$)     &   -0.4559    & -0.4482      &  -0.4599       &  -0.4020       &     -0.3643    \\
$D^a_{^{13}P_1}$~(GeV$^{-4}$)     &   -0.4160    & -1.2190      &  -2.2506       &  -4.1153       &     -6.3484     \\
\hline
$C_{^{33}P_1}$~(GeV$^{-4}$)       &    0.2907    &  0.1157      &  -0.0195       &  -0.3544       &     -0.5406    \\
$D^1_{^{33}P_1}$~(GeV$^{-6}$)     &    0.8531    &  2.0920      &   3.3395       &   5.6829       &      6.6607     \\
$C^a_{^{33}P_1}$~(GeV$^{-2}$)     &    1.0588    &  1.0661      &   1.0414       &   1.0537       &      1.0059    \\
$D^a_{^{33}P_1}$~(GeV$^{-4}$)     &    0.9332    &  1.7228      &   2.4337       &   3.4801       &      4.8833     \\
\hline
$C_{^{13}P_2}$~(GeV$^{-4}$)       &   -0.8469    & -0.9706      &  -1.0649       &  -1.2138       &     -1.2308    \\
$D^1_{^{13}P_2}$~(GeV$^{-6}$)     &    0.3952    &  0.2482      &  -0.3207       &  -1.5914       &     -3.1043     \\
$C^a_{^{13}P_2}$~(GeV$^{-2}$)     &    0.6605    &  0.7363      &   0.7913       &   0.8053       &      0.8707    \\
$D^a_{^{13}P_2}$~(GeV$^{-4}$)     &    1.8307    &  2.6057      &   3.7346       &   5.7949       &      7.7493     \\
\hline
$C_{^{33}P_2}$~(GeV$^{-4}$)       &    0.1264    & -0.1083      &  -0.2666       &  -0.3997       &     -0.4734    \\
$D^1_{^{33}P_2}$~(GeV$^{-6}$)     &    0.4147    &  0.3267      &   0.0475       &  -0.4968       &     -1.2914     \\
$C^a_{^{33}P_2}$~(GeV$^{-2}$)     &    0.6398    &  0.6177      &   0.6184       &   0.6079       &      0.6372    \\
$D^a_{^{33}P_2}$~(GeV$^{-4}$)     &    0.9179    &  1.8514      &   2.9132       &   4.5059       &      5.9872     \\
\hline
$D_{^{1}\epsilon_2}$~(GeV$^{-6}$)   &-0.4631     & -1.1534      &  -2.1726       &  -3.4707       &     -5.7668    \\
$D^a_{^{1}\epsilon_2}$~(GeV$^{-4}$) &-0.3377     & -0.4064      &  -0.3548       &  -0.4562       &     -0.1701    \\
\hline
$D_{^{3}\epsilon_2}$~(GeV$^{-6}$)   &0.1114      &  0.4043      &   0.8166       &   1.3557       &      2.0829    \\
$D^a_{^{3}\epsilon_2}$~(GeV$^{-4}$) &0.3512      &  0.3201      &   0.2310       &   0.1717       &      0.0662    \\
\hline
$D_{^{13}D_1}$~(GeV$^{-6}$)        &   -0.4469    & -1.4452      &  -2.5109       &  -4.0544       &     -5.5814    \\
$D^a_{^{13}D_1}$~(GeV$^-3$)        &    0.7330    &  0.8074      &   0.9080       &   1.0382       &      1.1982     \\
\hline
$D_{^{33}D_1}$~(GeV$^{-6}$)        &    0.6502    &  0.3119      &  -0.3121       &  -1.1332       &     -2.0036    \\
$D^a_{^{33}D_1}$~(GeV$^-3$)        &   -0.6839    & -0.8382      &  -0.9446       &  -1.0763       &     -1.1376     \\
\hline
$D_{^{11}D_2}$~(GeV$^{-6}$)       &    0.1535    &  0.2645      &   0.2236       &   0.0899       &     -0.2041    \\
$D^a_{^{11}D_2}$~(GeV$^{-3}$)     &    1.4617    &  1.6196      &   1.7957       &   1.9952       &      2.2209     \\
\hline
$D_{^{31}D_2}$~(GeV$^{-6}$)       &    1.1000    &  1.1738      &   1.1897       &    1.1904       &      1.1901    \\
$D^a_{^{31}D_2}$~(GeV$^{-3}$)     &    1.5223    &  1.6776      &   1.8608       &    2.0602       &      2.2818     \\
\hline
$D_{^{13}D_2}$~(GeV$^{-6}$)       &    0.0385    &  0.2183      &   0.5820       &   1.2472       &      2.2921    \\
$D^a_{^{11}D_2}$~(GeV$^{-3}$)     &    0.7547    &  0.8248      &   0.9059       &   0.9811       &      1.0547     \\
\hline
$D_{^{33}D_2}$~(GeV$^{-6}$)       &    0.8531    &  0.4431      &  -0.1465       &  -0.9415       &     -2.0262    \\
$D^a_{^{33}D_2}$~(GeV$^{-3}$)     &    1.0468    &  1.1852      &   1.3522       &   1.5445       &      1.7653     \\
\hline
$D_{^{13}D_3}$~(GeV$^{-6}$)       &   -1.9596    & -2.9471      &  -4.2752       &  -5.9646       &     -8.1915    \\
$D^a_{^{13}D_3}$~(GeV$^{-3}$)     &    1.0617    &  1.3561      &   1.6726       &   2.0161       &      2.4035     \\
\hline
$D_{^{33}D_3}$~(GeV$^{-6}$)       &    0.0403    & -0.5623      &  -1.2899       &  -2.1695       &     -3.2530    \\
$D^a_{^{33}D_3}$~(GeV$^{-3}$)     &    0.5867    &  0.6915      &   0.8157       &   0.9481       &      1.0945     \\
\hline
$D^a_{^{13}F_2}$~(GeV$^{-4}$) &   0.0275     &  0.0027      &  -0.2935       &  -0.5429       &     -1.6646    \\
\hline
$D^a_{^{33}F_2}$~(GeV$^{-4}$) &   1.5887     &  1.8414      &   2.0667       &   2.3631       &      2.6138   \\
\hline
$D^a_{^{11}F_3}$~(GeV$^{-4}$) &   1.4174     &  1.6372      &   1.8492       &   2.1002       &      2.3750    \\
\hline
$D^a_{^{31}F_3}$~(GeV$^{-4}$) &   0.8281     &  0.9437      &   1.0480       &   1.1818       &      1.3226    \\
\hline
$D^a_{^{33}F_3}$~(GeV$^{-4}$) &    0.7244    &  0.8419      &   0.9617       &   1.1091       &      1.3167    \\
\hline
$D^a_{^{13}F_4}$~(GeV$^{-4}$) &    1.4364    &  1.8230      &   2.0347       &   2.3687       &      2.6703    \\
\hline
\end{tabular}
\caption{\label{tab:cutoff1} Low-energy constants at N$^3$LO for different cutoffs.
The superscript $a$ indicates parameters that are related to the annihilation part,
cf. Eqs.~(17)-(\ref{ANN3}).
Note that all parameters are in units of $10^4$.
}
\end{center}
\end{table}

\begin{table}[htbp]
\begin{center}
\vspace{-1.5cm}
\begin{tabular}{|c||c|c|c|c|c|c|}
\hline\hline
\rule[-0.3cm]{0cm}{0.7cm}\multirow{2}{*}{\rule[-0.85cm]{0cm}{1.6cm}LEC}    & \multicolumn{3}{c|}{\rule{0.075cm}{0cm} R=0.9~fm \rule{0.075cm}{0cm}}    & \multicolumn{3}{c|}{\rule{0.075cm}{0cm} R=1.0~fm \rule{0.075cm}{0cm}}     \\
\cline{2-7}
\rule[-0.3cm]{0cm}{0.7cm} & \rule{0.3cm}{0cm} ~LO \rule{0.3cm}{0cm}    & \rule{0.3cm}{0cm} NLO\rule{0.3cm}{0cm}     &\rule{0.3cm}{0cm} ~N$^2$LO\rule{0.3cm}{0cm} & \rule{0.3cm}{0cm} ~LO \rule{0.3cm}{0cm}    & \rule{0.3cm}{0cm} NLO\rule{0.3cm}{0cm}     &\rule{0.3cm}{0cm} ~N$^2$LO\rule{0.3cm}{0cm}           \\
\hline\hline
$\tilde{C}_{^{11}S_0}$~(GeV$^{-2}$)   &   0.0200     & -0.0726      &  -0.0293     &   0.0150     & -0.0571      &  -0.0278         \\
$C_{^{11}S_0}$~(GeV$^{-4}$)           &   -          &  0.1624      &   0.1644     &   -          &  0.3056      &   0.2802           \\
$\tilde{C}^a_{^{11}S_0}$~(GeV$^{-1}$) &  -0.4500     & -0.4067      &  -0.4328     &  -0.4500     & -0.4252      &  -0.4373         \\
$C^a_{^{11}S_0}$~(GeV$^{-3}$)         &   -          & -0.9403      &  -0.6266     &   -          & -1.2876      &  -1.0742            \\
\hline
$\tilde{C}_{^{31}S_0}$~(GeV$^{-2}$)   &   -0.0075    & -0.0210      &   0.1218     &   0.0074     & -0.0161      &   0.0769         \\
$C_{^{31}S_0}$~(GeV$^{-4}$)           &   -          &  0.2930      &  -0.0128     &   -          &  0.3382      &   0.1738       \\
$\tilde{C}^a_{^{31}S_0}$~(GeV$^{-1}$) &    0.3547    & -0.4365      &  -0.4626     &   0.3859     & -0.4425      &  -0.4704        \\
$C^a_{^{31}S_0}$~(GeV$^{-3}$)         &   -          &  0.2317      &   0.8382     &   -          &  0.0029      &   0.4391         \\
\hline
$\tilde{C}_{^{13}S_1}$~(GeV$^{-2}$)   &   -0.1114    & -0.1719      &  -0.1076     &   -0.1052    & -0.2016      &  -0.1608       \\
$C_{^{13}S_1}$~(GeV$^{-4}$)           &    -         & -0.1932      &  -0.2336     &    -         & -0.2354      &  -0.3287        \\
$\tilde{C}^a_{^{13}S_1}$~(GeV$^{-1}$) &    0.3762    &  0.3471      &   0.3681     &    0.4154    &  0.4075      &   0.3943       \\
$C^a_{^{13}S_1}$~(GeV$^{-3}$)         &    -         &  0.9711      &   0.8904     &    -         &  1.5316      &   1.5600         \\
\hline
$\tilde{C}_{^{33}S_1}$~(GeV$^{-2}$)   &   -0.0500    & -0.1065      &   0.0112      &  -0.0116     & -0.0795      &  -0.0002        \\
$C_{^{33}S_1}$~(GeV$^{-4}$)           &    -         &  0.1539      &   0.2132      &   -          &  0.4228      &   0.3774        \\
$\tilde{C}^a_{^{33}S_1}$~(GeV$^{-1}$) &    0.4200    &  0.3577      &   0.4317      &   0.3250     &  0.3939      &   0.4240      \\
$C^a_{^{33}S_1}$~(GeV$^{-3}$)         &    -         &  1.5860      &   0.7752      &   -          &  1.5899      &   1.0812        \\
\hline
$C_{^{1}\epsilon_1}$~(GeV$^{-4}$)     &    -   & -0.2161      &  -0.1561       &   -     & -0.4420      &  -0.3025        \\
$C^a_{^{1}\epsilon_1}$~(GeV$^{-3}$)   & -      & -1.0121      &  -0.5084       &   -     & -1.1932      &  -0.8302       \\
\hline
$C_{^{3}\epsilon_1}$~(GeV$^{-4}$)     & -      &  0.1946      &   0.1926       &   -     &  0.2989      &   0.2950       \\
$C^a_{^{3}\epsilon_1}$~(GeV$^{-3}$)   & -      &  0.1793      &  -0.0675       &   -     &  0.1037      &  -0.0907      \\
\hline
$C^a_{^{13}D_1}$~(GeV$^{-3}$)          & -      &  0.0047      &  0.0061       &   -     & -0.0514      &  -0.6735       \\
\hline
$C^a_{^{33}D_1}$~(GeV$^{-3}$)          & -      &  0.7722      &  0.0008       &   -     &  0.8389      &   0.0096       \\
\hline
$C_{^{13}P_0}$~(GeV$^{-4}$)           &    -         & -1.4210      &  -0.6686      &   -     & -2.0451      &  -1.3926        \\
$C^a_{^{13}P_0}$~(GeV$^{-2}$)         &    -         & -0.7734      &  -0.7830      &   -     & -1.1055      &  -1.0913        \\
\hline
$C_{^{33}P_0}$~(GeV$^{-4}$)           &    -         & -0.9448      &  -0.6822       &   -     & -0.9678      &  -0.7343        \\
$C^a_{^{33}P_0}$~(GeV$^{-2}$)         &    -         &  0.7298      &   0.7532       &   -     &  0.8808      &   0.8977       \\
\hline
$C_{^{11}P_1}$~(GeV$^{-4}$)           &    -         &  0.2666      &   0.4396       &   -     &  0.3388      &   0.4880        \\
$C^a_{^{11}P_1}$~(GeV$^{-2}$)         &    -         & -0.8808      &  -0.9129       &   -     & -0.8863      &  -0.9032      \\
\hline
$C_{^{31}P_1}$~(GeV$^{-4}$)           &    -         &  0.0132      &   0.5938       &   -     &  0.0786      &   0.5748       \\
$C^a_{^{31}P_1}$~(GeV$^{-2}$)         &    -         & -0.8616      &  -0.9454       &   -     & -0.9090      &  -0.9541       \\
\hline
$C_{^{13}P_1}$~(GeV$^{-4}$)           &    -         & -0.5135      &  -0.1235        &   -     & -0.3593      &  -0.0221      \\
$C^a_{^{13}P_1}$~(GeV$^{-2}$)         &    -         & -0.5375      &  -0.5704        &   -     & -0.6041      &  -0.6258       \\
\hline
$C_{^{33}P_1}$~(GeV$^{-4}$)           &    -         & -0.0296      &   0.4602        &   -     & -0.0244      &   0.3859      \\
$C^a_{^{33}P_1}$~(GeV$^{-2}$)         &    -         &  0.8612      &   0.9263        &   -     &  0.9137      &   0.9481     \\
\hline
$C_{^{13}P_2}$~(GeV$^{-4}$)           &    -         & -0.9858      &  -0.4097        &   -     & -1.0905      &  -0.6203       \\
$C^a_{^{13}P_2}$~(GeV$^{-2}$)         &    -         & -0.8514      &  -0.9091        &   -     & -0.9919      &  -1.0219      \\
\hline
$C_{^{33}P_2}$~(GeV$^{-4}$)           &    -         & -0.5386      &  -0.1399        &   -     & -0.6099      &  -0.2712      \\
$C^a_{^{33}P_2}$~(GeV$^{-2}$)         &    -         &  0.6813      &   0.7159        &   -     &  0.7784      &   0.7992      \\
\hline
$C^a_{^{11}D_2}$~(GeV$^{-3}$)          & -      &  1.5335      &   1.5509       &   -     & 1.6924      &   1.7002        \\
\hline
$C^a_{^{31}D_2}$~(GeV$^{-3}$)          & -      &  1.5558      &   1.6436       &   -     & 1.7238      &   1.7697        \\
\hline
$C^a_{^{13}D_2}$~(GeV$^{-3}$)          & -      &  0.7062      &   0.7562       &   -     & 0.8087      &   0.8227       \\
\hline
$C^a_{^{33}D_2}$~(GeV$^{-3}$)          & -      &  1.1532      &   1.1551       &   -     & 1.3171      &   1.2937       \\
\hline
$C^a_{^{13}D_3}$~(GeV$^{-3}$)          & -      &  1.5684      &   1.4289       &   -     & 1.8667      &   1.7359       \\
\hline
$C^a_{^{33}D_3}$~(GeV$^{-3}$)          & -      &  0.7430      &   0.6654       &   -     & 0.8528      &   0.7784       \\
\hline\hline
\end{tabular}
\caption{\label{tab:order} Low-energy constants at LO, NLO, and N$^2$LO for the
cutoffs R=0.9~fm and 1.0~fm.
The superscript $a$ indicates parameters that are related to the annihilation part,
cf. Eqs.~(17)-(\ref{ANN3}).
Note that all parameters are in units of $10^4$.
}
\end{center}
\end{table}


\end{document}